\documentclass[12pt]{JHEP3}
\usepackage{mathrsfs}
\usepackage{amsmath,amssymb}
\usepackage{epsfig}
\usepackage{relsize}
\usepackage{bbm}                        
\usepackage{dcolumn}                    

\newcommand{\beq}{\begin{equation}}
\newcommand{\eeq}{\end{equation}}
\newcommand{\be}{\begin{equation}}
\newcommand{\ee}{\end{equation}}
\newcommand{\bea}{\begin{eqnarray}}
\newcommand{\eea}{\end{eqnarray}}
\newcommand{\ba}{\begin{array}}
\newcommand{\ea}{\end{array}}

\newcommand{\bra}[1]{\langle{#1}|}
\newcommand{\ket}[1]{|{#1}\rangle}

\newcommand{\nn}{\nonumber}

\newcommand{\ns} \normalsize
\newcommand{\kb}{\bar{\kappa}}

\newcommand{\PI}{\text{\Large{$\pi$}}}

\author{Per Sundin \\ \\
 {\it Humboldt-Universit\"at zu
Berlin, Institut f\"ur Physik,\\ Newtonstra{\ss}e 15, D-12489
Berlin, Germany}\\ \email{per.sundin@physik.hu-berlin.de}}
\abstract{We perform a detailed study of the type IIA superstring in AdS$_4 \times \mathbbm{CP_3}_3$. After introducing suitable bosonic light-cone and fermionic kappa worldsheet
gauges we derive the pure boson and fermion SU($2|2)\times$U(1) covariant light-cone Hamiltonian up to quartic order in fields. 

As a first application of our derivation we calculate energy shifts for string configurations in a closed fermionic subsector and successfully match these with a set of light-cone Bethe equations.

We then turn to investigate the mismatch between the degrees of freedom of scattering states and oscillatory string modes. Since only light string
modes
appear as fundamental Bethe roots in the scattering theory, the physical
role of the remaining $4_F+4_B$ massive oscillators is rather unclear. By continuing a
line of research initiated by Zarembo, we shed light on this question by
calculating quantum corrections for the
propagators of the bosonic massive fields. We show that, once loop corrections are incorporated,
 the massive coordinates dissolve in a continuum state of
two light particles.}

\title{On the worldsheet theory of the type IIA AdS$_4 \times \mathbbm{CP_3}_3$ superstring}
\preprint{HU-EP-09/41}
\begin{document}
\section{Introduction}
Recently strings on AdS$_4\times\mathbbm{CP_3}_3$ have enjoyed an increased
interest due to the AdS$_4$ / CFT$_3$ duality proposed in \cite{Schwarz:2004yj},
\cite{Aharony:2008ug}. The conjecture, nowadays dubbed
ABJM duality in the literature, states that
a three dimensional $\mathcal{N}=6$ and SU(N) Chern Simons theory living on the boundary of AdS$_4$
are in certain limits dual to type IIA string theory on AdS$_4 \times \mathbbm{CP_3}_3$. 

The duality exhibits many shared features with the well studied AdS$_5$ / CFT$_4$ correspondence, where perhaps the most striking similarity is the emergence of
integrable structures \cite{Nishioka:2008gz}, \cite{Gaiotto:2008cg}, \cite{Grignani:2008is}. On the gauge theory side, integrability
was demonstrated for the two loop Hamiltonian\footnote{The one loop piece
vanishes trivially.} in \cite{Minahan:2008hf}. Quickly after, the algebraic curve encoding all the classical solutions at strong and weak
coupling together with the all loop asymptotic Bethe equations were
put forward in \cite{Gromov:2008fy}, \cite{Gromov:2008qe}, \cite{Gromov:2008bz}. There after, and under the assumptions of a SU($2|2)\times$U(1) symmetry, the
exact S matrix were proposed in \cite{Ahn:2008aa}. Following these findings, a host of
various checks and higher order calculations have been performed \cite{Alday:2008ut},
\cite{McLoughlin:2008ms}, \cite{Krishnan:2008zs}, \cite{McLoughlin:2008he},
\cite{Kristjansen:2008ib}, \cite{Kalousios:2009ey}, \cite{Bak:2009mq}, \cite{Suzuki:2009sc}, \cite{Hamilton:2009iv}, \cite{Bak:2008cp}, \cite{Bak:2008vd}, \cite{Zwiebel:2009vb}, \cite{Minahan:2009te}, \cite{Spill:2008yr}, \cite{Grignani:2008is}, \cite{Ahn:2009zg}, \cite{Dimov:2009rd}, \cite{Schimpf:2009rk}.

That all this has been achieved with such a rapid progress is remarkable
since in both dualities the full dynamics can be constructed
from symmetry arguments alone. For ABJM, the symmetry group is OSP$(2,2|6)$,
which differs quite much from the well known PSU($2,2|4$) of AdS$_5$ / CFT$_4$.
Nevertheless, planar integrability, all loop asymptotic Bethe equations, SU($2|2)$ scattering and central extension
occur in similar ways in both dualities. 

In this paper we will perform a detailed study of the string theory side
of the ABJM correspondence. Starting from the symmetry group we derive the
Lagrangian in a super matrix notation utilizing an uniform light-cone gauge. 

As has been demonstrated by
Bykov in \cite{Bykov:2009jy}, the symmetry of the gauge fixed string reduces from OSP($2,2|6$) to
a centrally extended SU$(2|2)\times$U(1). This is rather similar
to the superstring in AdS$_5\times$S$^5$ which after gauge fixing have a
centrally extended SU$(2|2)^2$
algebra \cite{Beisert:2005tm}. Even though the gauge fixed subalgebras are rather similar, we find that the general structure
of the type IIA superstring is considerably more involved than its AdS$_5\times$S$^5$
cousin.

In order to extract any information from the Lagrangian we need to consider some sort of perturbative expansion. We will make use of a strong coupling expansion, or equivalently,  
an expansion in number of fields. Utilizing this expansion we derive the pure boson and fermion part of the light-cone Hamiltonian up to quartic
order in number of fields \cite{Callan:2003xr}, \cite{Callan:2004ev}, \cite{Callan:2004uv}, \cite{Frolov:2006cc}.

To avoid the rather severe complications of gauge fixing the worldsheet metric,
we work in a first order formalism. This has the upshot that the metric components
only enters as Lagrange multipliers. However, the theory exhibits higher order fermionic worldsheet time derivatives and to preserve a canonical Poisson structure
we need to shift the fermions in a appropriate way. Unfortunately, due
to the presence of cubic kinetic terms, this shift adds a 'self interacting' term which is very
hard to remove. Not only is the structure complicated, but it also introduces
corrections to the bosonic momentas. The way we approach this problem is
to only present the canonical Hamiltonian for pure boson / fermion fields. We do however present the full light-cone Hamiltonian, prior to the fermionic shift, in the appendix.

Having established the relevant parts of the first order theory, and under the assumption of normal ordering, we calculate
energy corrections to a
certain set of fermionic string states. Even though the general structure of relevant
parts of the Hamiltonian is rather involved, we find that the energy shifts
takes a remarkably simple form. This feature was also observed for the bosonic subsector
calculated in \cite{Sundin:2008vt} and seems to be a general feature of the
uniform light-cone gauge we imposed. We then match the energy shifts with
the predictions coming from a conjectured set of Bethe equations proposed
in \cite{Gromov:2008qe}, and rewritten in a light-cone language in \cite{Hentschel:2007xn} and \cite{Sundin:2008vt}. This is the first
calculation that explicitly probes the higher order fermionic sectors, and thus test the two body factorization implied by integrability, of
the AdS$_4$ / CFT$_3$ duality\footnote{Where we with higher order mean operators constituted of an arbitrary number of fermionic excitations.}.  

After this we turn to investigate the role of the massive modes of the theory.
At the quadratic level the string oscillators come in $4_F+4_B$
heavy and light modes respectively. From the point of view of the conjectured
exact scattering theory \cite{Beisert:2006ez}, the fundamental excitations in the S matrix are
the light modes, leaving us with a miss match between the degrees of freedom.

In \cite{Zarembo:2009au} Zarembo calculated the loop corrections for a massive bosonic mode.
There it was found that when quantum corrections are taken into account,
the analytic properties of the propagator changes. What happens is that the
pole gets shifted onto the branch cut and vanishes. Therefore
the heavy mode is not fundamental but rather a composite continuum state of two light particles.

We continue this line of research by showing that exactly the same thing
happens with the remaining massive bosons. Even though we do not calculate
it explicitly, we also provide some general arguments for why the same thing
should happen with the remaining massive fermionic coordinates.

The paper is organized as follows; We start out in section two by presenting
some general facts about the (super)matrix representation of the $\mathfrak{osp}(2,2|6)$
algebra. Then by making use of the $\mathbbm{Z}_4$ grading of the algebra, we construct
the exact string Lagrangian in a convenient kappa and light-cone gauge. In section three we expand the derived theory in a strong coupling
limit, equivalent to a near plane-wave expansion, to quartic order. We
find that the theory exhibits higher order time derivatives of the fermions,
and thus naively introduces a complicated Poisson structure. To tackle this
problem, we follow \cite{Frolov:2006cc} and introduce a fermionic shift with the property that
it removes the higher order kinetic terms. Sadly, this shift comes with the
price of adding additional cubic and quartic terms to the interacting Hamiltonian.
In section four we turn to a perturbative analysis of the string spectrum
by calculating energy shifts for fermionic states. These we then match
with a set of uniform light-cone Bethe equations, finding perfect agreement.
The last analysis we perform is to calculate loop diagrams for the bosonic
heavy modes in section six. We show that all the massive bosonic modes dissolve
into a two particle continuum, and therefore, do not appear as fundamental
excitations of the scattering theory.

We end the paper with a short summary and outlook together with several appendices
where notation and various computational details are presented.
\section{Type IIA superstring on AdS$_4\times \mathbbm{CP_3}_3$}
One of the most beautiful and effective ways to describe a physical theory
is through the use of its symmetries. An especially nice approach using
algebraic properties of a certain type of string configurations
has been developed
by Arutyunov and Frolov, see \cite{Arutyunov:2009ga} for a nice review. In the below we will apply this procedure
for a supersymmetric AdS$_4 \times \mathbbm{CP_3}_3$ string propagating on
the supergroup manifold \cite{Arutyunov:2008if} \cite{Stefanski:2008ik}
\bea 
\frac{\textrm{OSP}(2,2|6)}{\textrm{SO}(1,3)\times\textrm{U}(3)}.
\eea
A crucial ingredient is the existence
of a $\mathbbm{Z}_4$ grading of the symmetry algebra which allows for a construction
of the string Lagrangian directly from its graded components \cite{Berkovits:1999zq}.

To illustrate the procedure, we begin this section
by reviewing some basic facts of the super algebra $\mathfrak{osp}(2,2|6)$.
\subsection{Matrix realization of $\mathfrak{osp}(2,2|6)$}
The super Lie algebra $\mathfrak{osp}(2,2|6)$ can be represented by $10\times 10$ matrices of the form
\begin{displaymath}
M=
\left( \begin{array}{cc}
X_{4\times4} & \theta_{4\times 6} \\ 
\eta_{6\times 4} & Y_{6\times 6} \\
\end{array} \right)
\end{displaymath}
where $X$ and $Y$ are even matrices whereas $\theta$ and $\eta$ are Grassmannian
odd. 

To single out the algebra of interest, the matrices $M$ has to satisfy the
following reality and transposition rules,
\begin{displaymath}
M^{st}
\left( \begin{array}{cc}
C_4 & 0 \\ 
0 & \mathbbm{1}_{6\times6} \\
\end{array} \right)+\left( \begin{array}{cc}
C_4 & 0 \\ 
0 & \mathbbm{1}_{6\times6} \\
\end{array} \right)M=0 
\end{displaymath}
\begin{displaymath}
M^\dagger
\left( \begin{array}{cc}
\Gamma^0 & 0 \\ 
0 & -\mathbbm{1}_{6\times6} \\
\end{array} \right)+\left( \begin{array}{cc}
\Gamma^0 & 0 \\ 
0 & -\mathbbm{1}_{6\times6} \\
\end{array} \right)M=0
\end{displaymath}
where the charge conjugation matrix satisfies $C_4^2=-\mathbbm{1}_{4\times4}$
and $\Gamma_0$ is one of the AdS$_4$ $\Gamma$-matrices. In the first appendix
we collect all the various matrices encountered in this section. The super transpose is defined as
\begin{displaymath}
M^{st}=
\left( \begin{array}{cc}
X^t & -\eta^t \\ 
\theta^t & Y^t \\
\end{array} \right), 
\end{displaymath}
and the above reality and transposition rules imply
\bea 
X^t=-C_4\,X\,C_4^{-1}\,\qquad Y^t=-Y,\qquad \eta=-\theta^t\,C_4, \qquad
\theta^*=\Gamma^0\,C_4\,\theta.
\eea
The even $X$ and $Y$ block correspond to the bosonic isometry groups USP(2,2)
and SO(6) of AdS$_4$
and $\mathbbm{CP_3}_3$ respectively. The odd blocks are related by conjugation
and constitute 24 real spinor variables. The reality condition on the fermionic
block $\theta$ relates \footnote{Through out the paper we will denote conjugated
objects with bar.}
\bea 
\theta_{4,i}=\bar{\theta}_{1,i},\qquad \theta_{3,i}=-\bar{\theta}_{2,i}.
\eea
As advocated, the super algebra $\mathfrak{osp}(2,2|6)$ admits a $\mathbbm{Z}_4$
decomposition as
\bea 
M=M^{(0)}\oplus M^{(2)}\oplus M^{(1)}\oplus M^{(3)}.
\eea
We want to construct an inner automorphism such that its stationary point
coincides with $\mathfrak{so}(1,3)\oplus \mathfrak{u}(3)$. This can be done
by introducing two matrices $K_4$ and $K_6$
\begin{displaymath}
K_4=\left( \begin{array}{cccc}
0  & 1 & 0 & 0 \\
-1 & 0 & 0 & 0 \\
0  & 0 & 0 & 1 \\
0  & 0 & -1& 0 \\
\end{array} \right), \qquad
K_6=\left( \begin{array}{cccccc}
0  & 1 & 0 & 0 & 0 & 0\\
-1 & 0 & 0 & 0 & 0 & 0\\
0  & 0 & 0 & 1 & 0 & 0\\
0  & 0 & -1& 0 & 0 & 0\\
0  & 0 & 0  & 0 & 0 & 1\\
0  & 0 & 0  & 0 & -1 & 0\\
\end{array} \right)
\end{displaymath}
which satisfy $K_4^2=-\mathbbm{1}$ and $K_6^2=-\mathbbm{1}$. These
two matrices together with the charge conjugation matrix allows us to define
an automorphism as \cite{Arutyunov:2008if}
\begin{displaymath}
\Omega(M)=
\left( \begin{array}{cc}
K_4 C_4& 0 \\ 
0 & -K_6 \\
\end{array} \right)\,M\,\left( \begin{array}{cc}
K_4C_4 & 0 \\ 
0 & -K_6 \\
\end{array} \right)^{-1}=\Upsilon\,M\,\Upsilon^{-1},
\end{displaymath}
which can be used to construct the different $\mathbbm{Z}_4$ components
\bea 
M^{(k)}=\frac{1}{4}\Big(M+i^{3k}\Omega(M)+i^{2k}\Omega^2(M)+i^k\Omega^3(M)\Big),
\eea
where each component $M^{(k)}$ is an eigenstate of $\Omega$,
\bea 
\Omega(M^{(k)})=i^kM^{(k)}.
\eea
The stationary subalgebra, $M^{(0)}$, coincides with $\mathfrak{so}(1,3)\oplus \mathfrak{u}(3)$ which is the part of $\mathfrak{osp}(2,2|6)$ we want to divide
out. 

The orthogonal complement $M^{(2)}$ is spanned by matrices
satisfying $\Upsilon\,M\,\Upsilon^{-1}=-M$, which boils down to the conditions
\bea 
\{X,\Gamma^5\}=0,\qquad \{Y,K_6\}=0.
\eea
These two equations can be solved by 
\bea 
X=x_\mu\Gamma^\mu,\qquad Y=Y_i T_i,
\eea
where the first parameterize SO(3,2)/SO(1,3) and the second SO(6)/U(3). For
the exact form of the $\Gamma_\mu$ and $T_i$ generators, please consult the
appendix.

With this we have established a good parameterization of $\mathfrak{osp}(2,2|6)$. In the next section we will construct the full
string Lagrangian from this.
\subsection{Group parameterization and string Lagrangian}
There are many ways to parameterize OSP(2,2$|$6) and they
are all related through non linear field transformations. In this paper we will
use a particulary suitable representation that allows us to fix the bosonic
and fermionic worldsheet symmetries in a convenient way \cite{Frolov:2006cc}. 

As starting point we introduce the following group element of OSP(2,2$|$6),
\bea
\label{GE}
G=\Lambda(x^+,x^-)\,f(\eta)\,G_t,
\eea
where the different components are given by
\bea \nn
\Lambda(x^+,x^-)=\exp{\frac{i}{2}(x^+\Sigma_++x^-\Sigma_-)}, \qquad G_t=G_y\,G_{AdS}\,G_{CP}, \qquad f(\eta)=\eta+\sqrt{1+\eta^2}.\eea
The $x^\pm=\phi\pm t$ are a light-cone pair constituted of the time and angle coordinate of AdS$_4$ and $\mathbbm{CP_3}_3$ and $\Sigma_\pm$ is the corresponding
basis element, $\Sigma_\pm=\pm \Gamma_0\oplus -i\,T_6$. The fermionic matrix $\eta$,
entering in $f(\eta)$, is in principle just the odd part of $M$. The transverse
bosonic degrees of freedom are described by $G_t$,
\begin{displaymath}
G_t=
\left( \begin{array}{cc}
G_{AdS}& 0 \\ 
0 & G_y\,G_{CP} \\
\end{array} \right).
\end{displaymath}
The AdS$_4$ part is parameterized by three transverse coordinates, $z_i$
\bea 
&& G_{Ads}=\frac{\mathbbm{1}+\frac{i}{2}z_i\Gamma^i}{\sqrt{1-\frac{z_i^2}{4}}}.
\eea
and the $G_y$ element is described by a single real coordinate, $y$, of the $\mathbbm{CP_3}_3$,
\bea 
G_y=e^{y\,T_5},
\eea
which is a function of $\cos(y)$ and $\sin(y)$. For the upcoming perturbative analysis it is convenient to relabel the trigonometric functions as
\bea \nn 
\sin(y)\rightarrow \frac{1}{2}y, \qquad \cos(y)\rightarrow \sqrt{1-\frac{1}{4}y^2}.
\eea
The last component of $G_t$ is parameterized by two complex coordinates $\omega_i$
(and its conjugate $\bar{\omega}_i$)
\bea
&& G_{CP}= \\ \nn 
&& \mathbbm{1}+\frac{1}{\sqrt{1+\frac{1}{4}|w|^2}}\big(W+\bar{W}\big)+4\,\frac{\sqrt{1+\frac{1}{4}|w|^2}-1}{|w|^2\sqrt{1+\frac{1}{4}|w|^2}}\big(W\cdot\bar{W}+\bar{W}\cdot
W\big),
\eea
where $W=\frac{1}{2}\omega_i\,\tau_i$ and $|w|^2=\omega_i\,\bar{\omega}_i$.

Using these parameterizations, we can construct a flat current in $\mathfrak{osp}(2,2|6)$
as
\bea 
\mathcal{A}=\mathcal{A}^{(0)}\oplus \mathcal{A}^{(2)}\oplus \mathcal{A}^{(1)}\oplus
\mathcal{A}^{(3)}=-G^{-1}\,dG,
\eea
which components are used to construct the string Lagrangian,
\bea 
\label{secondorderL}
\mathscr{L}=-\frac{g}{2}\int d\sigma\,Str\Big(\gamma^{\alpha\beta}\mathcal{A}^{(2)}_\alpha\, \mathcal{A}^{(2)}_\beta+\kappa\,\epsilon^{\alpha\beta}\mathcal{A}^{(1)}_\alpha\,\mathcal{A}^{(3)}_\beta\Big).
\eea
Throughout the paper we will use greek letters for worldsheet indices.
The string length parameter is denoted $\sigma$ and takes values $\sigma \in [-L,L]$. The variable $\kappa$ in front of the WZ term
is demanded by supersymmetry to satisfy\footnote{It is also related to parity
invariance of $\sigma$. Sending $\sigma \rightarrow -\sigma$ induces a sign
change of $\kappa$.} $\kappa^2=1$. The $\gamma^{\alpha\beta}$
tensor is the Weyl invariant combination of the worldsheet metric with determinant $det\,\gamma^{\alpha\beta}=-1$. Finally, the model is characterized by the string coupling, $g\sim \frac{R^2}{\alpha'}$
with $R$ the radius of the AdS space. This is the only free parameter of
the theory and later we will expand the theory in a limit with $g$ taken
as large.

Since our aim is to perform a perturbative expansion of the above Lagrangian the
gauge fixing procedure gets considerably simplified if we introduce a auxiliary field
$\PI$ which allows us to rewrite the Lagrangian (\ref{secondorderL}) as \cite{Frolov:2006cc}
\bea
\label{firstorderL}
&& \mathscr{L}= \\ \nn 
&& -g \int d\sigma \,Str\Big(\PI\,\mathcal{A}_0+\frac{\kappa}{2}\,\epsilon^{\alpha \beta}\,\mathcal{A}_\alpha^{(1)}\,\mathcal{A}_\beta^{(3)}-\frac{1}{2\gamma^{00}}\big(\PI^2+(\mathcal{A}_1^{(2)})^2\big)+\frac{\gamma^{01}}{\gamma^{00}}\,\PI\,\mathcal{A}_1^{(2)}\Big).
\eea
Using the equations of motion for $\PI$ one can easily show that this Lagrangian
is classically equivalent to (\ref{secondorderL}). The metric components
of $\gamma^{\alpha\beta}$ enter as Lagrange multipliers giving rise to the
two constraints
\bea
\label{constraints} 
Str\,\PI^2+Str\big(\mathcal{A}^{(2)}_1\big)^2=0, \qquad Str\,\PI\,\mathcal{A}_1^{(2)}=0.
\eea
Loosely speaking the solution of the first constraint give
the gauge fixed string Hamiltonian while the second allow us to solve for one of the unphysical light-cone coordinates. 

The auxiliary field $\PI$ allows for a basis decomposition with respect to $\mathfrak{osp}(2,2|6)$,
\bea 
\PI=\PI_+\Sigma_++\PI_-\Sigma_-+\PI_t,
\eea 
where
\begin{displaymath}
\PI_t=
\left( \begin{array}{cc}
\PI_i^{(z)}\,\Gamma_i& 0 \\ 
0 & \PI^{(y)}\,T_5+\PI_i^{(\omega)}\,\tau_i+\bar{\PI}^{(\bar{\omega})}_i\,\bar{\tau}_i
\end{array} \right)
\end{displaymath}
Note that from this basis decomposition, one see that $Str\,\PI\,\mathcal{A}^{(2)}=Str\,\PI\,\mathcal{A}^{even}=Str\,\PI\,\mathcal{A}$
which we used in (\ref{firstorderL}). 

One can think about the field $\PI$ as the matrix version of a first order
formalism. By introducing this field we effectively get rid of the worldsheet
metric which make the process of bosonic gauge fixing considerably simpler.
However, it is important to understand that the components of $\PI$ does
not directly correspond to the conjugate momentas of the bosonic fields.
In order to obtain the physical Hamiltonian, one have to solve for these components
and use the solutions in the Lagrangian (\ref{firstorderL}). We will discuss
this point in more detail in the next section. 
\subsection{Gauge fixing and field content}
The Lagrangian (\ref{secondorderL}) and (\ref{firstorderL}) are invariant
under two dimensional diffeomorphisms, Weyl scalings and fermionic kappa symmetry where the latter
is a local worldsheet symmetry with odd transformation parameter.

The
bosonic symmetries are used to fix a uniform light-cone gauge as
\bea 
x^+=\sigma^0=\tau, \qquad p_+=\textrm{Constant},
\eea
which has the important consequence that the string coupling, $g$,
becomes related to the length of the string, $g \sim L$ \cite{Arutyunov:2006gs}.

The model contains 24 real fermions whereas supersymmetry demands that the
number of fermionic and bosonic excitations should be equal.
At first glance, this looks like a problem since common lore has is that kappa
symmetry removes half of the fermions, which in our case would leave us with
to few fermions for supersymmetry to be manifest. However, as it turns out, the
kappa symmetry for strings in AdS$_4 \times \mathbbm{CP_3}_3$ is partial and
only allows for eight real fermions to be removed \cite{Arutyunov:2008if}. Therefore the kappa fixed
model has equal number of fermionic and bosonic excitations.

There are many ways to impose the kappa symmetry. In this paper we will
use an especially convenient gauge introduced by Bykov which is compatible
with the bosonic part of the subgroup that commutes with the gauge-fixed string Hamiltonian\footnote{For another covariant kappa gauge, see \cite{Dukalski:2009pr}.} \cite{Bykov:2009jy}.

It can be shown that the light-cone Hamiltonian is proportional to $Str\,Q\,\Sigma_+$,
where $Q$ is the Noether charge associated with the global OSP(2,2$|$6) symmetry
\cite{Arutyunov:2006ak}.
A specific symmetry generator can be expressed as a linear combination of
$Q$ traced over various basis element of $\mathfrak{osp}(2,2|6)$.
The commutator between two charges, say $Q_1=Str \,Q \,\mathcal{M}_1$
and $Q_1=Str \,Q \,\mathcal{M}_2$, is given by
\bea 
[Q_1,Q_2]_\pm \sim Str\,Q\,[\mathcal{M}_1,\mathcal{M}_2]_\pm,
\eea
where the $\pm$ is $+$ only if both charges are odd. It is easy
to see that the subalgebra that commutes with the light-cone Hamiltonian
is given by matrices of the form
\bea 
\{\mathcal{M}\in\mathfrak{osp}(2,2|6)\,,\, [\Sigma_+,\mathcal{M}]=0\}=\mathfrak{su}(2|2)\oplus
\mathfrak{u}(1).
\eea
The bosonic part of this subalgebra is $\mathfrak{su}(2)_{AdS}\oplus \mathfrak{su}(2)_{CP} \oplus \mathfrak{u}(1)_{CP}$, where the subscript denotes which space the
isometry originates from. 
In a matrix notation, elements in the bosonic subalgebra
takes the form
\begin{displaymath}
\mathfrak{g}_B=
\left( \begin{array}{ccc}
\mathfrak{su}(2)_{AdS}\vert_{4\times 4} & 0 & 0\\
0& \mathfrak{u}(1)_{CP}\vert_{2\times 2} & 0 \\ 
0 & 0 & \mathfrak{su}(2)_{CP}\vert_{4\times 4} \\
\end{array} \right)
\end{displaymath}
A important fact is that all elements in $\mathfrak{g}_B$ also commute with $\Sigma_-$. This has the important consequence
that these transformations
only act on the transverse part of the
group elment,
\bea
[\Sigma_\pm,\mathfrak{g}_B]=0 \rightarrow [\mathfrak{g}_B,\Lambda(x^+,x^-)]=0.
\eea
So, if we let $e^{\mathfrak{g}_B}$ act on $G$ we find
\bea 
e^{\mathfrak{g}_B}\cdot G=\Lambda(x^+,x^-)\,g_{B}\,f(\eta)\,g^{-1}_{B}\,g_{B}\,G_t\,g^{-1}_{B}\cdot
g_c,
\eea 
where $g_c$ is a irrelevant compensating transformation from the stabilizer group, SO(1,3)$\times$U(3).
From this we see that the bosons and the fermions are in the adjoint representation
of $G^{B}$=SU(2)$\times$SU(2)$\times$U(1). 

As was explained in \cite{Bykov:2009jy}, a kappa gauge that transform covariantly under $G^B$
can be constructed by first enforcing
\bea 
\label{kappa}
\theta_{1,5}=i\,\theta_{1,4},\quad \theta_{1,6}=i\,\theta_{1,3}, \quad \theta_{2,5}=i\,\theta_{2,4},
\quad \theta_{2,6}=i\,\theta_{2,3},
\eea
which removes four complex fermions and thus leave us with a total of sixteen
real ones as desired\footnote{One can also think about the kappa gauge in the following way; if we anticommute a generic, non kappa gauge fixed odd matrix, with
$\Sigma_+$, one find that the resulting object has the form of a kappa gauge fixed
matrix. In one sense this can be seen as a defining property of the gauge. This is very similar to the kappa gauge imposed in \cite{Frolov:2006cc} where the
gauge fixing was defined through a commutation relation between
a light-cone basis element and $\eta$.}. As it stands, the gauge (\ref{kappa}), does not transform covariantly under
the bosonic symmetries. However, if we augment the gauge with the following
linear combinations of the spinor components\footnote{Note that the fermions denoted with $\kappa^\pm$ has no relation
with the constant $\kappa$ in front of the WZ term in the Lagrangian. Also
note that the
$\pm$ denotes U(1) charge and should not be confused as sign of
the SU(2) index.}
\bea 
\label{fermlabel}
\nn 
&& \theta_{1,1}=\kappa^{+\,1}-\kappa_{+\,2}, \quad \theta_{1,2}=-i(\kappa^{+\,1}+\kappa_{-\,2}),
\quad \theta_{2,1}=\kappa^{+\,2}+\kappa_{-\,1}, \quad \theta_{2,2}=-i(\kappa^{+\,2}-\kappa_{-\,1}),
\\ 
&& \theta_{1,3}=\frac{1}{2}(s^1_{\dot{1}}-s^1_{\dot{2}}), \quad \theta_{1,4}=-\frac{i}{2}(s^1_{\dot{1}}+s^1_{\dot{2}}),
\quad \theta_{2,3}=\frac{1}{2}(s^2_{\dot{1}}-s^2_{\dot{2}}), \quad \theta_{2,4}=-\frac{i}{2}(s^2_{\dot{1}}+s^2_{\dot{2}}),
\eea
then the new variables transform under $G^B$ as
\bea 
\kappa^{+,a} \rightarrow e^{i\alpha}\,g^a_b \,\kappa^{+\,b}, \quad
\kappa_{-\,a} \rightarrow e^{-i\alpha}\,g_a^b \,\kappa_{-\,b}, \quad
s^a_{\dot{b}}\rightarrow g^a_b\,g^{\dot{a}}_{\dot{b}}\,s^b_{\dot{a}},
\eea
where $g^a_b \in$ SU$(2)_{AdS}$, $g^{\dot{a}}_{\dot{b}}\in$ SU(2)$_{CP}$ and $e^{\pm i \phi}\in$ U(1). Thus, in our notation, undotted
indices correspond to the SU(2) from the AdS space and dotted ones correspond
to the SU(2) from $\mathbbm{CP_3}_3$. In this notation it becomes clear that
we have two set of spinors, $\kappa^\pm$, with opposite U(1) charge transforming under the AdS SU(2)\footnote{The spinor transforming with negative U(1)
is in the conjugate representation of the SU(2) from AdS, hence the lower index.}. There is also a spinor, $s^a_{\dot{b}}$, uncharged under the
U(1) but in a bifundamental representation of the two SU(2)'s.

We should also classify how the bosonic fields transform. Clearly, the $z_i$ coordinates only transform under the SU(2) from the AdS space. The singlet $y$ does not transform
at all, neither under any SU(2) or the U(1). The only bosonic fields charged under
the U(1) are the complex $\omega_i$ and $\bar{\omega}_i$ which also transform under the SU(2)
of $\mathbbm{CP_3}_3$. A convenient index notation is
\bea 
\omega_i \rightarrow \omega_{\dot{a}}, \qquad \bar{\omega}_i \rightarrow
\bar{\omega}^{\dot{a}},
\eea
where lower index has the plus charge of the U(1) and vice versa.

Under conjugation, all indices changes place
\bea 
\label{indices}
&&(\kappa^{+\,a})^\dagger=\bar{\kappa}_{+\,a}=\epsilon_{ab}\,\bar{\kappa}^{+\,b},
\quad (\kappa_{-\,a})^\dagger=\bar{\kappa}^{-\,a}=\epsilon^{ab}\,\bar{\kappa}_{-\,b},
\\ \nn 
&& (s^a_{\dot{b}})^\dagger=\bar{s}_a^{\dot{b}}=\epsilon^{\dot{b}\dot{a}}\epsilon_{ab}\,\bar{s}^b_{\dot{a}},
\quad (\omega_{\dot{a}})^\dagger=\bar{\omega}^{\dot{a}}=\epsilon^{\dot{a}\dot{b}}\,\omega_{\dot{b}}, \quad \epsilon_{ab}\,\epsilon^{bc}=\delta^c_a,
\quad \epsilon_{\dot{a}\dot{b}}\,\epsilon^{\dot{b}\dot{c}}=\delta^{\dot{c}}_{\dot{a}},
\eea
where we also introduced epsilon tensors to raise and lower indices, with the convention $\epsilon_{01}=1=-\epsilon^{01}$. It is
convenient to let the 
$\pm$, denoting U(1) charge of the unconjugated spinors, travel with the SU(2)
index. This imply that all lower $\pm$ have negative U(1) while upper have
positive. 
\subsection{Light-cone Lagrangian and Hamiltonian}
Having imposed the bosonic and fermionic gauges, we are in position to start
extracting physical quantities from the string Lagrangian (\ref{firstorderL}).
The most natural object to study is of course the string Hamiltonian. In the
light-cone formalism it is given by minus the conjugate momenta of $x^+$
and it enters the Lagrangian in the natural way
\bea \nn 
\mathscr{L}=p_m \,\dot{x}^m+p_-+\textrm{Fermions}, \qquad m\in\{i,y,\dot{a}\}.
\eea
The Hamiltonian, $-p_-$, is a function of the physical
fields and the auxiliary field $\PI$. The auxiliary field does not directly correspond
to the momentum variables of the bosonic fields. Rather, each component of $\PI$ can
be expressed, and solved for, in terms of them. To extract the light-cone
Hamiltonian in terms of physical fields we will proceed below as follows; for all but the $\PI_-$ component, we use the conjugate
momentas to solve for the components, that is
\bea \nn 
\frac{\delta \mathscr{L}}{\delta \dot{x}_M}=p_M=f(\textrm{fields}\neq \PI_M,\PI_M)\rightarrow
\PI_M=\tilde{f}(\textrm{fields}\neq p_M,p_M), \qquad M\neq -.
\eea
Doing this for the transverse momenta shows that\footnote{As can be seen, the complex components mix within each other and one might
be tempted to shift the fields so this complication disappears. However,
as it turns out this mixing enters only at quartic order in number of fields
so for the upcoming perturbative analysis this mixing is irrelevant.}
\bea 
\label{pitrans}
&& \PI^{(z)}_i=\frac{2i\,p_i^{(z)}}{4+z_i^2}, \quad \PI^{(y)}=\frac{4\,p_y}{8+y^2-\omega_{\dot{a}}\,\bar{\omega}^{\dot{a}}},
\\ \nn 
&&  \PI_{\dot{1}}^{(\omega)}=\frac{8\,p_{\dot{1}}+\omega_{\dot{1}}\,\bar{\omega}^{\dot{2}}\,\PI_{\dot{2}}^{(\omega)}}{8-\omega_{\dot{1}}\,\bar{\omega}^{\dot{1}}-\omega_{\dot{a}}\,\bar{\omega}^{\dot{a}}},
\quad 
\PI_{\dot{2}}^{(\omega)}=\frac{8\,p_{\dot{2}}+\omega_{\dot{2}}\,\bar{\omega}^{\dot{1}}\,\PI_{\dot{1}}^{(\omega)}}{8-\omega_{\dot{2}}\,\bar{\omega}^{\dot{2}}-\omega_{\dot{a}}\,\bar{\omega}^{\dot{a}}},
\eea
which more or less by definition satisfy
\bea 
\label{boskin}
Str \, \PI \,G_t^{-1}\,\dot{G}_t=p_m\,\dot{x}^m,
\eea
where $m$ runs over transverse indices. 

The expressions for $\PI_\pm$ are considerably more complicated
and for these components we will only present the corresponding matrix equations\footnote{However, their quadratic part is needed to determine the upcoming fermionic shift, so these parts we present in (\ref{piquadratic}).}.
To obtain $\PI_+$ we solve for $p_+$ in a similar way as we did above, then use
this solution in the quadratic constraint (\ref{constraints}) to solve for $\PI_-$,
\bea
\label{pi}
&& \PI_+=-\PI_-\,\frac{Str\,\Sigma_-\,G_-}{Str\,\Sigma_+\,G_-}+\frac{1}{Str\,\Sigma_+\,G_-}\Big(\mathbbm{p}_+-Str\,\PI_t\,G_-\Big), \\ \nn
&& \PI_-=\frac{\mathbbm{p}_+-Str\,\PI_t\,G_-}{2\,Str\,\Sigma_-\,G_-}
\Big\{1\pm\sqrt{1-\frac{\big(Str\,\Sigma_-\,G_-\big)\big(Str\,\Sigma_+\,G_-\big)\Big(Str\,\PI_t^2+Str\big(\mathcal{A}_1^2\big)^2\Big)}{4\,\big(\mathbbm{p}_+-Str\,\PI_t\,G_-\big)^2}}\Big\}\\ \nn
&& = \frac{\big(Str\,\Sigma_+\,G_-\big)\Big(Str\,\PI_t^2+Str\big(\mathcal{A}_1^2\big)^2\Big)}{16\,\Big(\mathbbm{p}_+-Str\,\PI_t\,G_-\Big)}+...
\eea
where we introduced the short hand notation $G_-$ for the even part of
\bea \nn 
\frac{i}{2}\,G_t^{-1}\,\big(f^{-1}(\eta) \,\Sigma_- \, f(\eta)\big)\,G_t
\eea
and $\mathbbm{p}_+$ is
\bea
&& \mathbbm{p}_+=p_+-p_+^{WZ}= \\ \nn
&& p_+-\kappa\frac{i}{2}\,Str\Big\{G_t^{-1}\Big(\frac{i}{2}\sqrt{1+\eta^2}\,\Sigma_-\,\eta-\frac{i}{2}\,\eta\,\Sigma_-\,\sqrt{1+\eta^2}\Big)G_t
\,\Upsilon\,\mathcal{A}^{Odd}_1\,\Upsilon^{-1}\Big\},
\eea
where the last part is the contribution to $p_+$ coming from the WZ term
and $\mathcal{A}^{Odd}=\mathcal{A}^{(1)}+\mathcal{A}^{(3)}$.

The light-cone Hamiltonian is given by
\bea
\label{lcham}
&& -\mathcal{H}=p_-=\frac{\delta\mathscr{L}}{\delta \dot{x}^+}= \\ \nn
&& \frac{i}{2}\,Str\,\PI\,G_t^{-1}\big(\Sigma_+-\eta\,\Sigma_+\,\eta+\sqrt{1+\eta^2}\,\Sigma_+\,\sqrt{1+\eta^2}\big) \,G_t \\ \nn
&& -\kappa\frac{i}{2}\,Str\Big\{G_t^{-1}\Big(\frac{i}{2}\sqrt{1+\eta^2}\,\Sigma_+\,\eta-\frac{i}{2}\,\eta\,\Sigma_+\,\sqrt{1+\eta^2}\Big)G_t
\times \\ \nn
&& \Upsilon\,G_t^{-1}\Big(\big(\frac{i}{2}\sqrt{1+\eta^2}\,\Sigma_-\,\eta-\frac{i}{2}\,\eta\,\Sigma_-\,\sqrt{1+\eta^2}\big)\,x'^-+\sqrt{1+\eta^2}\,\eta'-\eta\,\partial_1 \sqrt{1+\eta^2}\Big)G_t\,\Upsilon^{-1}\Big\}.
\eea
As it stands, the expression above is very involved. To be able to extract anything
useful from it one need to consider various simplifying limits, which will
be the main topic of the next section. 

Combining everything we have so far, we can write the string Lagrangian as
\bea
\label{exactL}
&& \mathscr{L}= \\ \nn 
&& p_+\,\dot{x}^- +p_m\,\dot{x}^m+p_-
+Str\,\PI\,G_t^{-1}\Big(-\eta\,\dot{\eta}+\sqrt{1+\eta^2}\,\partial_0\sqrt{1+\eta^2}\Big)G_t
\\ \nn 
&&+\frac{i}{2}\kappa\,Str\,G_t^{-1}\Big(\sqrt{1+\eta^2}\,\partial_0\eta-\eta\,\partial_0\sqrt{1+\eta^2}\Big)
\,G_t\,\Upsilon\,\mathcal{A}_1^{Odd}\,\Upsilon^{-1}.
\eea
Together with the solutions for $\PI$ and the expression for $p_-$ in (\ref{lcham})
this is the $exact$ gauge fixed string Lagrangian for the AdS$_4\times\mathbbm{CP_3}_3$
superstring. It will be the starting point for a perturbative analysis
in the next section. However, it should be clear that the terms involving
time derivatives of the fermions will have terms beyond quadratic order.
This severely complicates the quantization procedure since we get a very
involved Poisson structure for the fermionic quantities, see \cite{Alday:2005jm} for an example. Luckily, one can side pass this complication
by performing a shift of the fermions in such a way that the higher order
kinetic terms vanish. This has the advantage of a canonical Poisson structure
but with the cost of additional terms in the light-cone Hamiltonian.
\section{Strong coupling expansion}
To be able to extract anything useful from (\ref{lcham}) we have to consider
some sort of perturbative expansion. The standard way to proceed is to
boost, spin or deform the string in some way or another. In this paper we will
expand around a point like string configuration moving on a null geodesic. Or equivalently, a plane wave expansion \cite{Berenstein:2003gb}. In
practise the limit boils down to the following expansion scheme\footnote{This
is essentially the same expansion scheme as for the string in \cite{Frolov:2006cc} and \cite{Sundin:2008vt} with the effective BMN coupling $\tilde{\lambda}$ put to unity.}
\bea 
\label{expansionscheme}
g\rightarrow \infty, \qquad x_m\rightarrow \frac{x_m}{\sqrt{g}}, \qquad p_m\rightarrow \frac{p_m}{\sqrt{g}}, \qquad \eta\rightarrow
\frac{\eta}{\sqrt{g}},
\eea
which becomes an expansion in number of fields \cite{Frolov:2006cc}.

An important physical consequence of the above limit is that the string length,
which was proportional to $g$, becomes infinite. The worldsheet of a closed
string has the topology of a cylinder so taking the $g\rightarrow \infty$
limit means that the string decompactifies. It becomes a infinite plane. In
terms of string Bethe equations and asymptotic configurations, this fact
has far reaching consequences, see \cite{Arutyunov:2009ga} and references therein. 
\subsection{Leading order}
It is a good idea to start out the perturbative analysis by fixing some of the
constants we encountered so far. First of all, from now on we will fix\footnote{Once
again we stress that the $\kappa$ here has nothing to do with the two fermions
$\kappa^\pm$.}
\bea 
p_+=1 \qquad \kappa=1.
\eea
What we choose to do with our parameter space is of course arbitrary and the physics we want to extract is totally independent of numerical conventions.
However, the choices
above are very convenient in terms of notation. Having factors of $\kappa$
and $p_+$ in the expressions makes things which are, and especially will
become, complicated more involved than necessary. 

It is also desirable to have the Lagrangian in such a form that the field
expansions becomes as simple as possible. To achieve this we rescale the string length parameter as
$\sigma\rightarrow 2 \sigma$ and send\footnote{This
is equivalent to defining the fermionic part of the group element as $f(\eta)=\sqrt{1-\eta^2}+i\,\eta$.} $\eta \rightarrow i\,\eta$. Taking this into consideration, and taking
the limit (\ref{expansionscheme}) of (\ref{exactL}) gives the leading order
quadratic Lagrangian
\bea
\label{L2real}
&& \frac{1}{2}\mathscr{L} = p_i\,\dot{z}_i+p_y\,\dot{y}+\dot{w}_{\dot{a}}\,\bar{p}^{\dot{a}}+\dot{\bar{\omega}}^{\dot{a}}\,p_{\dot{a}}+i\bar{s}_a^{\dot{b}}\,\dot{s}^a_{\dot{b}}+i\bar{\kappa}_{+\,a}\,\dot{\kappa}^{+\,a}
+i\bar{\kappa}^{-\,a}\,\dot{\kappa}_{-\,a}\\ \nn
&& -p_i^2-4\bar{p}^{\dot{a}}\,p_{\dot{a}}-p_y^2-\frac{1}{4}\big(y^2+z_i^2+\frac{1}{4}\bar{\omega}^{\dot{a}}\,\omega_{\dot{a}}\big)-\frac{1}{4}\big(
z_i'^2+y'^2+\bar{\omega}'^{\dot{a}}\,\omega'_{\dot{a}}\big)\\ \nn
&& -\bar{s}_a^{\dot{b}}\,s^a_{\dot{b}}-\frac{1}{2}\big(\bar{\kappa}_{+\,a}\,\kappa^{+\,a}+\bar{\kappa}^{-\,a}\,\kappa_{-\,a}\big)-i\big(\kappa_{-\,a}\,\kappa'^{+\,a}+\bar{\kappa}_{+\,a}\,\bar{\kappa}'^{-\,a}\big)\\ \nn
&& -\frac{i}{2}\big(s^a_{\dot{b}}\,(s')_a^{\dot{b}}+\bar{s}_a^{\dot{b}}\,(\bar{s}')^a_{\dot{b}}\big).
\eea
From this we find that the fields come in heavy and light multiplets,
\bea \nn 
\textbf{M}=1;\qquad \{s^a_{\dot{b}}\,,\,z_i\,,\,y\}\qquad \textbf{M}=\frac{1}{2}; \qquad \{\kappa^{+\,a}\,,\,\kappa_{-\,a}\,,\,\omega_{\dot{a}}\,,\,\bar{\omega}^{\dot{a}}\}.
\eea
This $4_{\frac{1}{2}}+4_{1}$ split of the masses is a novel feature for the
AdS$_4 \times \mathbbm{CP_3}_3$ string. In the last section of this paper we
will calculate loop corrections to propagators for the massive modes.
There it will be argued that the heavy excitations can be viewed as composite
states of light modes. For now though we view them as single excitations.

Note that we all through out the paper work with phase space variables. The gauge fixing procedure is vastly simplified through the use of the auxiliary $\PI$ field since it allowed us to eliminate the dependence of the worldsheet metric. The auxiliary field is expressed in terms of the two unknown $\PI_\pm$ components and the transverse momentum variables. If we desired, we could after the gauge fixing procedure is completed, express the momentum variables in terms of velocities resulting in a different, but completely equivalent, formulation of the theory, see \cite{Callan:2003xr} and \cite{Callan:2004ev}. However, as in for example \cite{Frolov:2006cc} and \cite{Sundin:2008vt}, we find it convenient to stick with the phase space formulation. Also, the parameterization of the group element that we use is especially suitable for a Hamiltonian analysis since the transverse coordinates of the auxilitary field $\PI$, in (\ref{pitrans}), do not depend on any fermionic quantities. 

We can tidy up the notation a bit further by making the quadratic 2-d Lorentz symmetry
manifest. First we introduce, $\gamma^0=\sigma_3$ and $\gamma^1=-i\sigma_2$, which obeys $\{\gamma^\alpha,\gamma^\beta\}=2\eta^{\alpha\beta}$ with $(+,-)$ convention. We then combine the fermions into two spinors as
\begin{displaymath}
\Psi=
\left( \begin{array}{c}
\kappa^{+\,a}  \\
\bar{\kappa}^{-\,a}
\end{array} \right), \quad \bar{\Psi}=\Psi^\dagger\,\gamma^0=
\left( \begin{array}{cc}
\kb_{+\,a} & ,-\kappa_{-\,a} 
\end{array} \right), \quad \chi=
\left( \begin{array}{c}
s^a_{\dot{b}} \\
\bar{s}^a_{\dot{b}}
\end{array} \right), \quad \bar{\chi}=\chi^\dagger\,\gamma^0=
\left( \begin{array}{cc}
\bar{s}^{\dot{b}}_a & ,-s_a^{\dot{b}} 
\end{array} \right).
\end{displaymath}
Then the quadratic Lagrangian can be written as
\bea
\label{quadraticL}
&&\frac{1}{2}\mathscr{L}= p_i\,\dot{z}_i+p_y\,\dot{y}+\dot{w}_{\dot{a}}\,\bar{p}^{\dot{a}}+\dot{\bar{\omega}}^{\dot{a}}\,p_{\dot{a}}
\\ \nn 
&& -p_i^2-4\bar{p}^{\dot{a}}\,p_{\dot{a}}-p_y^2-\frac{1}{4}\big(y^2+z_i^2+\frac{1}{4}\bar{\omega}^{\dot{a}}\,\omega_{\dot{a}}\big)-\frac{1}{4}\big(
z_i'^2+y'^2+\bar{\omega}'^{\dot{a}}\,\omega'_{\dot{a}}\big)
\\
\nn 
&&+i\bar{\Psi}\,\gamma^\alpha\,\partial_\alpha\,\Psi+\frac{i}{2}\bar{\chi}\,\gamma^\alpha
\,\partial_\alpha\,\chi-\frac{1}{2}\bar{\Psi}\,\Psi-\frac{1}{2}\bar{\chi}\,\chi.
\eea
Anticipating the quantization procedure we expand the fields in Fourier coefficients
as
\bea \nn
&& \omega_{\dot{a}}=\frac{1}{\sqrt{2\,\pi}}\int\,dp\,\frac{1}{\sqrt{\omega_p}}\Big(a^{\dot{a}}\,e^{ip\sigma}+\bar{b}^{\dot{a}}\,e^{-ip\sigma}\Big),
\quad p_{\dot{a}}=\frac{i}{\sqrt{2\,\pi}}\int\,dp\,\frac{\sqrt{\omega_p}}{4}\Big(\bar{b}^{\dot{a}}\,e^{-ip\sigma}-a^{\dot{a}}\,e^{ip\sigma}\Big),
\\ \nn 
&& y=\frac{1}{\sqrt{2\,\pi}}\int\,dp\,\frac{1}{\sqrt{2\Omega_p}}\Big(y\,e^{ip\sigma}+\bar{y}\,e^{-ip\sigma}\Big),
\quad
p_y=\frac{1}{2}\frac{i}{\sqrt{2\,\pi}}\int\,dp\,\sqrt{\frac{\Omega_p}{2}}\Big(\bar{y}\,e^{-ip\sigma}-y\,e^{ip\sigma}\Big),
\\ \nn 
&& z_i=\frac{1}{\sqrt{2\,\pi}}\int\,dp\,\frac{1}{\sqrt{2\Omega_p}}\Big(z_i\,e^{ip\sigma}+\bar{z}_i\,e^{-ip\sigma}\Big),
\quad
p_i=\frac{1}{2}\frac{i}{\sqrt{2\,\pi}}\int\,dp\,\sqrt{\frac{\Omega_p}{2}}\Big(\bar{z}_i\,e^{-ip\sigma}-z_i\,e^{ip\sigma}\Big),
\\ \nn 
&& s^a_{\dot{b}}=\frac{1}{\sqrt{2\pi}}\int dp \frac{1}{\sqrt{2\Omega_p}}\Big(F_p\,\chi^a_{\dot{b}}\,e^{ip\sigma}-H_p\,\bar{\chi}^a_{\dot{b}}\,e^{-ip\sigma}\Big),
\\ \nn 
&& \kappa^{+\,a}=\frac{1}{\sqrt{2\pi}}\int dp \frac{1}{\sqrt{2\omega_p}}\Big(f_p\,c^a\,e^{ip\sigma}-h_p\,\bar{d}^{\,a}\,e^{-ip\sigma}\Big),
\\ \nn 
&& \kappa_{-\,a}=\frac{1}{\sqrt{2\pi}}\int dp \frac{1}{\sqrt{2\omega_p}}\Big(f_p\,d_a\,e^{ip\sigma}-h_p\,\bar{c}_a\,e^{-ip\sigma}\Big),
\eea 
and obvious ones for conjugated fields. The frequencies and the fermionic wave functions are given by,
\bea 
&& \omega_p=\sqrt{\frac{1}{4}+p^2}, \quad f_p=\sqrt{\frac{\omega_p+\frac{1}{2}}{2}},
\quad h_p=\frac{p}{2f_p}, \\ \nn 
&& \Omega_p=\sqrt{1+p^2}, \quad F_p=\sqrt{\frac{\Omega_p+1}{2}}, \quad H_p=\frac{p}{2F_p},
\eea
where the wave functions satisfy the following important identities, 
\bea \nn 
f_p^2+h_p^2=\omega_p, \quad f_p^2-h_p^2=\frac{1}{2},\quad F_p^2+H_p^2=\Omega_p,
\quad F_p^2-H_p^2=1.
\eea
If we now plug the field expansion into (\ref{L2real}) and integrate over
$\sigma$, we find
\bea 
&& L=\int dp\,\Big(i\big(\bar{b}^{\dot{b}}\,\dot{b}_{\dot{b}}+\bar{a}_{\dot{b}}\,\dot{a}^{\dot{b}}+\bar{y}\,\dot{y}+\bar{z}_i\,\dot{z}_i+\bar{\chi}_a^{\dot{b}}\,\dot{\chi}^a_{\dot{b}}+\bar{c}_a\,\dot{c}^a+\bar{d}^a\,\dot{d}_a\big)\\
\nn 
&&-\omega_p\big(\bar{b}^{\dot{b}}\,b_{\dot{b}}+\bar{a}_{\dot{b}}\,a^{\dot{b}}+\bar{c}_{\dot{b}}\,c^a+\bar{d}^a\,d_a\big)-\Omega_p\big(\bar{y}\,y+\bar{z}_i\,z_i+\bar{\chi}_a^{\dot{b}}\,\chi^a_{\dot{b}}\big)\Big).
\eea
We also need to consider the second constraint in (\ref{constraints}) which
give rise to 
\bea 
\label{levelmatching}
\mathcal{V}=\int \, dp \,p\Big(\bar{b}^{\dot{b}}\,b_{\dot{b}}+\bar{a}_{\dot{b}}\,a^{\dot{b}}+\bar{y}\,y+\bar{z}_i\,z_i+\bar{c}_{\dot{b}}\,c^a
+\bar{d}^a\,d_a+\bar{\chi}_a^{\dot{b}}\,\chi^a_{\dot{b}}\Big).
\eea
Which is the so called level matching constraint enforcing that the sum of
all mode numbers has to vanish for physical states. In the quantum theory
this will be promoted to an operator whose action on a physical state should
project
to zero.

Promoting the oscillators to operators is now down by imposing the equal
time (anti)commutators
\bea
\label{commutationrelations}
&& [a(p,\tau)^{\dot{a}},\bar{a}(p',\tau)_{\dot{b}}]=2\pi\,\delta_{\dot{b}}^{\dot{a}}\,\delta(p-p'),
\quad [b(p,\tau)_{\dot{a}},\bar{b}(p',\tau)^{\dot{b}}]=2\pi\,\delta^{\dot{b}}_{\dot{a}}\,\delta(p-p')
\\ \nn 
&& [y(p,\tau),\bar{y}(p',\tau)]=2\pi\,\delta(p-p'), \quad [z_i(p,\tau),\bar{z}_j(p',\tau)]=2\pi\,\delta_{ij}\delta(p-p'),
\\ \nn
&& \{c^a(p,\tau),\bar{c}_b(p',\tau)\}=\{d_b(p,\tau),\bar{d}^a(p',\tau)\}=2\pi\,\delta^a_b\,\textsl{}\delta(p-p'),
\\ \nn 
&& \{\chi^a_{\dot{a}}(p,\tau),\bar{\chi}_{b}^{\dot{b}}(p',\tau)\}=2\pi\,\delta^a_b\,\delta^{\dot{a}}_{\dot{b}}\,\delta(p-p').
\eea
With this we have established the quadratic Lagrangian, including field expansions
and commutation relations. We would now like to proceed to the higher order contributions from (\ref{lcham}). However, before extracting the  sub leading
terms in
the light-cone Hamiltonian, we have to take care of the higher order kinetic
fermions. If these were to be included then the anti commutation relations
in (\ref{commutationrelations}) would receive higher order corrections. In
the next section we will describe how this complication can (partially) be avoided by
a appropriate shift of the fermions.
\subsection{Canonical fermions}
The focus of this section be will the piece of (\ref{exactL}) that contains kinetic fermionic terms,
\bea 
\label{kineticfermL}
&&  \mathscr{L}^{\eta}_{Kinetic}= \\ \nn
&& \frac{1}{2}Str\,\PI\,G_t^{-1}\Big([\dot{\eta},\eta]+\frac{1}{4}[\eta^2,\{\dot{\eta},\eta\}]\Big)G_t \\ \nn
&& -\frac{i}{2}\kappa\,Str\,G_t^{-1}\Big(\dot{\eta}-\frac{1}{2}\eta\,\dot{\eta}\,\eta\Big)
\,G_t\,\Upsilon\,G_t^{-1}\big(\frac{i}{2}\,[\Sigma_-,\eta]\,x'^-+\eta'-\frac{1}{2}\eta\,\eta'\,\eta\big)G_t\,\Upsilon^{-1}+\mathcal{O}(\eta^6),
\eea
from which it is clear that the anti commutation relations in (\ref{commutationrelations})
will receive higher order contributions. In principle this is not a fundamental
problem and it can be solved explicitly by a careful analysis of the Poisson
structure, see for example \cite{Alday:2005jm}. However, from a calculational point of view,
it is rather cumbersome to deal with non trivial commutation relations. For
that reason we will try to avoid the problem by performing a shift of the
fermionic coordinates\footnote{For a similar but much simpler discussion, see \cite{Frolov:2006cc}.}.

By using the cyclicity of the super trace and the form of $\PI_+$, we can
write\footnote{$\PI_+$ is the only component of the auxiliary field
which has a constant leading order term.}
\bea
\label{shiftshort}
\mathscr{L}^\eta_{Kinetic}=\frac{i}{4}\,Str\,\Sigma_+\,\dot{\eta}\,\eta+Str\,\dot{\eta}\,\widetilde{\Phi}(x_m,p_m,\eta),
\eea
where $\widetilde{\Phi}(x_m,p_m,\eta)$ is a complicated fermionic matrix,
presented in (\ref{formoffermshift}),
that can be deduced from (\ref{kineticfermL}). It starts at quadratic order
in number of fields and for the analysis at hand we have to know it up to
cubic order\footnote{The observant reader might notice that (\ref{kineticfermL})
also has a second quadratic piece $\sim Str\, \dot{\eta}\,\Upsilon\,\eta'\,\Upsilon^{-1}$.
This term is, however, a total derivative and can be neglected.}

We will now show that most of the higher order terms can be removed by shifting the fermions in an appropriate way. First we introduce a, so
far arbitrary, function $\Phi(x_m,p_m,\eta)$. Since we are to expand the
Hamiltonian up to quartic order, we need this function to third order in
number of fields. To simplify the notations we split up $\Phi(x_m,p_m,\eta)$
in number of fields and leave the bosonic dependence implicit, $\Phi(x_m,p_m,\eta)=\Phi_2(\eta)+\Phi_3(\eta)$.
The idea is now to shift the fermionic matrix as
\bea 
\label{fermshifteta}
\eta \rightarrow \eta+\Phi(\eta).
\eea 
Performing
the shift in (\ref{shiftshort}) and writing,
$\widetilde{\Phi}(x_m,p_m,\eta)=\widetilde{\Phi}_2(\eta)+\widetilde{\Phi}_3(\eta)$, we find
\bea
\label{fermshift}
&&  \mathscr{L}^{\eta}_{Kinetic}=\\ \nn 
&& \frac{i}{4}\,Str\,\Sigma_+\,\dot{\eta}\,\eta+ Str\,\dot{\eta}\,\big(\widetilde{\Phi}_2(\eta)+\widetilde{\Phi}_3(\eta)\big)+\frac{i}{4}\,Str\,\dot{\eta}[\Phi_2(\eta)+\Phi_3(\eta),\Sigma_+] \\ \nn
&& +Str\,\dot{\eta}\,\widetilde{\Phi}_2(\eta\rightarrow\Phi_2)+Str\,\dot{\Phi}_2(\eta)\,\widetilde{\Phi}_2(\eta)
+\frac{i}{4}\,Str\,\Sigma_+\,\dot{\Phi}_2(\eta)\,\Phi_2(\eta),
\eea
where $\widetilde{\Phi}_2(\eta\rightarrow\Phi_2)$ is a cubic contribution
from $\widetilde{\Phi}$ with $\Phi_2$ as argument.

To proceed, we need to find the form of $\Phi$. We do this by recalling that a general kappa gauge fixed fermionic element, which we again call $\eta$, can be written as a commutator, $\eta=[\Sigma_+,\chi]$ for some arbitrary, non kappa gauge fixed, fermionic matrix $\chi$. This means that a term of the form $Str\,\dot{\eta}\,\widetilde{\Phi}$, for arbitrary fermionic $\widetilde{\Phi}$, can be written $Str \, \dot{\chi}\,[\Sigma_+,\widetilde{\Phi}]$. This imply that for $\Phi$ to remove the higher order terms, it should satisfy the matrix
equation
\bea
[\Sigma_+,[\Phi,\Sigma_+]]+[\Sigma_+,\widetilde{\Phi}]=0.
\eea
Some trial and error shows that a solution for $\Phi$ in terms of $\widetilde{\Phi}$
is
\begin{displaymath}
\Phi=
\left( \begin{array}{cc}
\mathbbm{1}_{6\times 6} & 0 \\
0 & \frac{1}{4}\,\mathbbm{1}_{4\times 4} \\
\end{array} \right)\,[\Sigma_+,\widetilde{\Phi}]\,
\left( \begin{array}{cc}
\mathbbm{1}_{6\times 6} & 0 \\
0 & \frac{1}{4}\,\mathbbm{1}_{4\times 4} \\
\end{array} \right)=\Gamma\,[\Sigma_+,\widetilde{\Phi}]\,\Gamma,
\end{displaymath}
which allows us to remove the $Str\,\dot{\eta}\,\widetilde{\Phi}$ terms in (\ref{fermshift}) by choosing,
\bea
\label{phi}
\Phi=-4i\,\Gamma\,[\Sigma_+,\widetilde{\Phi}_2+\widetilde{\Phi}_2(\eta\rightarrow\Phi_2)+\widetilde{\Phi}_3]\, \Gamma.
\eea
This leaves us with
\bea
&&  \mathscr{L}^{\eta}_{Kin}=\frac{i}{4}\,Str\,\Sigma_+\,\dot{\eta}\,\eta+Str\,\dot{\Phi}_2\,\widetilde{\Phi}_2
+\frac{i}{4}\,Str\,\Sigma_+\,\dot{\Phi}_2\,\Phi_2,
\eea
which can be rewritten using (\ref{phi}) to
\bea 
\label{kinferm1}
\mathscr{L}^\eta_{Kin}=\frac{i}{4}\,Str\,\Sigma_+\,\dot{\eta}\,\eta+\frac{1}{2}Str\,\dot{\Phi}_2\,\widetilde{\Phi}_2.
\eea
The last expression is unfortunately rather involved. It is of quartic order
in number of fields and introduce additional time derivatives of the bosonic
fields since
\bea \nn
\widetilde{\Phi}_2=\frac{1}{2}\big(\frac{i}{4}[\eta,[G_t^{1},\Sigma_+]]+[\eta,\PI_t^1]\big)-\frac{i}{2}\big([G_t^1,\Upsilon]\,\eta'\,\Upsilon^{-1}+\Upsilon\,\eta'\,[G_t^1,\Upsilon^{-1}]\big),
\eea
where $G_t^1$ and $\PI_t^1$ are the pieces of $G_t$ and $\PI_t$ linear in
fields. To remove the additional fermionic kinetic terms induced by the shift,
one needs to isolate the $\dot{\eta}$ terms from (\ref{kinferm1}) and introduce
a second shift, say $\hat{\Phi}_3$, with the property $\frac{i}{4}Str\,\dot{\eta}[\hat{\Phi}_3,\Sigma_+]=-\frac{1}{2}Str\,\Phi_2\,\widetilde{\Phi}_2\vert_{\dot{\eta}}$,
where the notation is meant to imply the $\dot{\eta}$ dependent
part of $Str \,\Phi_2\,\widetilde{\Phi}_2$.
However, this means that the $\dot{\eta}$ independent part contains time
derivatives of the bosonic fields, so we find corrections to the transverse
part of $\PI$ in (\ref{pitrans}). Needless to say, this analysis
becomes rather involved. Not only will the additional fermionic shift, $\hat{\Phi}_3$,
complicate things further, but the additional momentum terms also give rise
to complications since they will have a quadratic fermionic
dependence\footnote{One could try to change the form of the OSP$(2,2|6)$
group element as $G=\Lambda\,G_t\,f(\eta)$ which simplifies the fermionic
kinetic term with the price of fermionic dependence in the bosonic conjugate
momentas from start. However, pushing through with the analysis one finds that in the
end the complications are more or less the same and the fermionic shift is still
very involved.}.

We will tackle this problem by simply ignoring it. Or, to be more precise,
we assume that the $\hat{\Phi}_3$ shift is performed but do not determine
the form of it, nor the additional momentum terms, allowing us to maintain
the canonical Poisson structure for the fermions. The reason we can do this
is because $Str\,\Phi_2\,\widetilde{\Phi}_2$ contains two fermions and two
bosons, which implies that all additional terms, both from the shift and
from $\PI_t$, will end up in the mixing
part of the shifted Hamiltonian, $\mathcal{H}_{BF}$. This is acceptable since
this part is not needed for the upcoming analysis.

However, a nice feature of the shift is that the $x'^-$ dependence will cancel between the shifted and the original quartic Hamiltonian\footnote{This is
also true for the shifted $\mathcal{H}_{BF}$ part. The additional contributions
from the complicated $Str\,\Phi_2\,\widetilde{\Phi}_2$ does introduce
any additional $x^-$ terms.}. Another nice consequence  of the shift is that it removes
all fermionic non $\sigma$ derivative terms from the relevant parts of the
Hamiltonian. This is important
since the point particle dynamics should be fully encoded in the quadratic
fluctuations. 

To summarize what we have done; We introduced a fermionic shift $\Phi$, which
can be expressed in terms of $\widetilde{\Phi}$, with the property that it
removes all higher order fermionic derivative terms. However, due to the
presence of cubic terms in the Lagrangian, the shift adds a 'self interaction'
term of the form $Str \,\Phi_2\,\widetilde{\Phi}_2$. This term is not only
complicated, but it also alters the transverse part of the auxiliary field
$\PI$. Instead of determining this term explicitly, we simply assume
the shift is performed, which guarantees a canonical Poisson structure. This is
equivalent to put $Str\,\Phi_2\,\widetilde{\Phi}_2$ to zero by hand and accept that we can not determine the mixing part, $\mathcal{H}_{BF}$, of the shifted Hamiltonian. It is a bit surprising that the fermions are of such a complicated nature. For the AdS$_5\times$S$^5$ string the corresponding shift actually simplified the resulting theory, while here it has the opposite effect. Perhaps it is related to the coset construction we use which is not as rigorous as the AdS$_5$ string, see \cite{Gomis:2008jt} and \cite{Grassi:2009yj} for a related discussion.

What we can determine though is the shifted part of the Hamiltonian containing
only bosons and fermions. This we will do in the next section. In the appendix
we also present the full unshifted Hamiltonian, which together with the full
form of the fermionic shift allows one to determine the shifted mixing Hamiltonian.
\subsection{Higher order Hamiltonian}
Having established the relevant form of the fermionic shift we are now in position
to derive the Hamiltonian (\ref{lcham}) to quartic order in fields. The way
to do this is a straight forward, albeit somewhat tedious, multi step process. First we use the solution
for $\PI$ in (\ref{lcham}), impose the shift (\ref{fermshifteta}) and expand
to quartic order. It should be obvious
that due to the complexity of both the Hamiltonian and the shift, it is very
desirable to use some sort of computer program that can handle symbolic manipulations\footnote{For
this paper we made use of Mathematica version 7 together with the package
\cite{jeremy}.}.

Pushing through with the calculation one find that the Hamiltonian has cubic
next to leading order terms. This is another novel feature compared to the AdS$_5\times$S$^5$ string which subleading terms start at quartic order.

Before we present our findings we would like to
introduce yet another convenient notation,
\bea 
&& Z^a_b=\sum_i\,z_i\sigma^a_{i,b}, \qquad Z^2=\frac{1}{2}Tr\,Z^a_b\,Z^b_c=\sum_i\,z_i^2\\
\nn 
&&
P^a_{z,b}=\sum_i\,p_{i}\sigma^a_{i,b},\qquad P_z^2=\frac{1}{2}Tr\,P^a_{z,b}\,P^b_{z,c}=\sum_i\,p_i^2,
\eea
where the Pauli matrices transform as $\sigma \rightarrow g\,\sigma\,g^t$
under the AdS SU(2). 

With all this, we are now in position to extract the full Hamiltonian. Starting out with the subleading cubic part, we find
\bea 
\label{H3twospinor}
&& \sqrt{g}\,\mathcal{H}_3= \\ \nn 
&& \big(\bar{\Psi}_a\,\Psi^b\big)'Z^a_b+i\big(\bar{\Psi}\,\gamma^1\,\Psi'-\bar{\Psi}'\,\gamma^1\,\Psi\big)_a^b\,(Z')^a_b-2i\big(\bar{\Psi}'\,\gamma^0\,\Psi-\bar{\Psi}\,\gamma^0\,\Psi'\big)_a^b\,P_{z,b}^a\\
\nn 
&& +2\Big(\big(\bar{\chi}_{a\beta}\,\gamma^1\,\Psi'^a-\bar{\chi}'_{ab}\,\gamma^1\,\Psi^a\big)\bar{p}^\beta+\big(\bar{\Psi}_a\,\gamma^1\,\chi'^{a\,\dot{b}}-\bar{\Psi}'_a\,\gamma^1\,\chi^{a\,\dot{b}}\big)\,p_\beta\Big)
+\frac{i}{4}\Big(\big(\bar{\chi}_{a\,\dot{b}}\,\gamma^1\,\gamma^0\,\Psi^a\big)'\bar{\omega}^{\dot{b}}
\\ \nn 
&&+\big(\bar{\Psi}_a\,\gamma^0\,\gamma^1\,\chi^{a\,\dot{b}}\big)'\omega_{\dot{b}}\Big)+\frac{1}{2}\big(\bar{\chi}_{a\,\dot{b}}\,\gamma^0\,\Psi'^a-\bar{\chi}'_{a\,\dot{b}}\,\gamma^0\,\Psi^a\big)\bar{\omega}'^{\dot{b}}
+\frac{1}{2}\big(\bar{\Psi}_a\,\gamma^0\,\chi'^{a\,\dot{b}}-\bar{\Psi}'_a\,\gamma^0\,\chi^{a\,\dot{b}}\big)\omega'_{\dot{b}}
\\ \nn 
&& +i\,y\,\big(\bar{p}^{\dot{b}}\,\omega_{\dot{b}}-p_{\dot{b}}\,\bar{\omega}^{\dot{b}}\big).
\eea
A nice feature of the coordinate system we use is that the massive singlet
do not mix with any of the fermionic coordinates. Let us also remark that
the fermionic shift (\ref{fermshifteta}) induces additional terms already
here in the cubic Hamiltonian. 

We will split up the quartic Hamiltonian according to its bosonic / fermionic field
content $g\,\mathcal{H}_4=\mathcal{H}_{BB}+\mathcal{H}_{BF}+\mathcal{H}_{FF}$.
For the pure bosonic contribution, we find
\bea 
\label{Hbb}
&& \frac{g}{2}\,\mathcal{H}_{BB}=\\ \nn 
&&\frac{1}{4}Z^2\,Z'^2-\frac{3}{4}\,p_y^2\,y^2+\frac{1}{16}y^4-\frac{1}{16}y^2\,y'^2-\frac{1}{16}\bar{\omega}^{\dot{a}}\,\bar{\omega}'^{\dot{b}}\,\omega_{\dot{b}}\,\omega'_{\dot{a}}
-\frac{3}{32}\bar{\omega}^{\dot{a}}\,\bar{\omega}'^{\dot{b}}\,\omega_{\dot{a}}\,\omega'_{\dot{b}}
\\ \nn 
&& 
-\frac{1}{128}\bar{\omega}^{\dot{a}}\,\bar{\omega}^{\dot{b}}\,\omega_{\dot{a}}\,\omega_{\dot{b}}
+\frac{1}{2}\bar{p}^{\dot{a}}\,\bar{\omega}^{\dot{b}}\,p_{\dot{a}}\,\omega_{\dot{b}}+\bar{p}^{\dot{a}}\,\bar{\omega}^{\dot{b}}\,p_{\dot{b}}\,\omega_{\dot{a}}
-\frac{1}{8}\bar{\omega}'^{\dot{a}}\,\omega'_{\dot{a}}\,y^2 -\frac{3}{32}\bar{\omega}^{\dot{a}}\,\omega_{\dot{a}}\,y'^2
\\
\nn 
&&
-2\,\bar{p}^{\dot{a}}\,p_{\dot{a}}\,y^2+\frac{1}{8}p_y^2\,\bar{\omega}^{\dot{a}}\,\omega_{\dot{a}}
-\frac{1}{2}y^2\,P_z^2-\frac{1}{8}\bar{\omega}^{\dot{a}}\,\omega_{\dot{a}}\,P_z^2
+2\,\bar{p}^{\dot{a}}\,p_{\dot{a}}\,Z^2+\frac{1}{8}\,\bar{\omega}'^{\dot{a}}\,\omega'_{\dot{a}}\,Z^2
\\ \nn 
&&-\frac{1}{32}\,\bar{\omega}^{\dot{a}}\,\omega_{\dot{a}}\,Z'^2 +\frac{1}{8}y'^2\,Z^2+\frac{1}{2}p_y^2\,Z^2-\frac{1}{8}y^2\,Z'^2,
\eea 
which, for another more complicated coordinate system, was first calculated in \cite{Sundin:2008vt}. 

Next we turn to the purely fermionic part which is given by\footnote{The
expression is not simplified by using the two spinor notation so we choose
to present it with the $s^a_{\dot{b}}$ and $\kappa^\pm$ terms explicit.},
\bea 
\label{Hff}
&& g\,\mathcal{H}_{FF}=-i\big(\kappa_{-\,a}\,\kb_{+\,b}\,\kappa^{+\,a}\,\kappa'^{+\,b} +\kappa_{-\,a}\,\kappa'_{-\,b}\,\kappa^{+\,a}\,\kb^{-\,b}\big)-\frac{i}{2}\big(\kappa_{-\,a}\,\kb_{+\,b}\,\kappa'^{+\,a}\,\kappa^{+\,b}\\ \nn
&& 
+\kappa_{-\,a}\,\kappa'_{-\,b}\,\kb^{-\,a}\,\kappa^{+\,b}  +\kappa_{-\,a}\,\kb'_{+\,b}\,\kb^{-\,a}\,\kb^{-\,b}
+\kb_{+\,a}\,\kb_{+\,b}\,\kappa^{+\,a}\,\kb'^{-\,b}\big) +\big(\kappa_{-\,a}\,\kb_{+\,b}\,\kb'^{-\,a}\,\kappa'^{+\,b}\\ \nn
&&
+\kappa_{-\,a}\,\kappa_{-\,b}\,\kappa'^{+\,a}\,\kappa'^{+\,b}+\kb_{+\,a}\,\kb_{+\,b}\,\kb'^{-\,a}\,\kb'^{-\,b}\big)-\frac{5}{2}\big(\kappa_{-\,a}\,\kappa_{-\,b}\,\kb'^{-\,a}\,\kb'^{-\,b} +\kb_{+\,a}\,\kb_{+\,b}\,\kappa'^{+\,a}\,\kappa'^{+\,b}\big)
 \\ \nn
&&   -3\,\kappa_{-\,a}\,\kb'_{+\,b}\,\kb'^{-\,a}\,\kappa^{+\,b}-4\big(\kappa_{-\,a}\,\kb_{+\,b}\,\kappa'^{+\,a}\,\kb'^{-\,b}
-\kappa_{-\,a}\,\kappa'_{-\,b}\,\kappa'^{+\,a}\,\kappa^{+\,b}  +\kappa_{-\,a}\,\kb'_{+\,b}\,\kappa^{+\,a}\,\kb'^{-\,b}
\\ \nn
&&
-\kappa_{-\,a}\,\kappa'_{-\,b}\,\kb^{-\,a}\,\kb'^{-\,b}-\kb_{+\,a}\,\kb'_{+\,b}\,\kappa^{+\,a}\,\kappa'^{+\,b} +\kappa_{-\,a}\kb'_{+\,b}\,\kappa'^{+\,a}\,\kb^{-\,b}+\kb_{+\,a}\,\kappa'_{-\,b}\,\kb'^{-\,a}\,\kappa^{+\,b}
\big)\\ \nn &&+5\big(\kappa_{-\,a}\,\kappa'_{-\,b}\,\kb'^{-\,a}\,\kb^{-\,b}+\kb_{+\,a}\,\kb'_{+\,b}\,\kappa'^{+\,a}\,\kappa^{+\,b}\big) +6\,\kappa_{-\,a}\,\kappa'_{-\,b}\,\kappa^{+\,a}\,\kappa'^{+\,b}-2\big(\kappa_{-\,a}\,\kb'_{+\,b}\,\kb^{-\,a}\,\kappa'^{+\,b}
\\ \nn 
&& +\kb_{+\,a}\,\kappa'_{-\,b}\,\kappa^{+\,a}\,\kb'^{-\,b}\big)-\frac{1}{2}(\bar{s}')^{\dot{a}}_a\,(\bar{s}')^{\dot{b}}_b\,s^{a}_{\dot{b}}\,s^{b}_{\dot{a}}
+\frac{1}{2}s^{a}_{\dot{a}}\,s^{b}_{\dot{b}}\,(s')^{c,\dot{b}}\,s'^{d\,\dot{a}}\big(\epsilon_{a\,b}\,\epsilon_{c\,d}-\epsilon_{a\,d}\,\epsilon_{c\,b}\big) \\ \nn
&& 
-\frac{i}{4}\big(\kappa_{-\,a}\,\kappa'^{+\,a}\,\bar{s}_b^{\dot{a}}\,s^b_{\dot{a}}-\kappa_{-\,a}\,\kappa'^{+\,b}\,\bar{s}_a^{\dot{b}}\,s^a_{\dot{a}}  -\kappa'_{-\,c}\,\kappa^{+\,a}\,\bar{s}_a^{\dot{a}}\,s^c_{\dot{a}}\big)  -\frac{1}{2}\Big(\,\kb'_{+\,c}\,\kappa^{+\,c}\,\bar{s}_a^{\dot{a}}\,(s')^a_{\dot{a}}-\kappa'_{-\,c}\,\kb'^{-\,c}\,\bar{s}_a^{\dot{a}}\,s^a_{\dot{a}}\\
 \nn 
 &&+\kappa'_{-\,c}\,\kb'^{-\,a}\,\bar{s}_a^{\dot{a}}\,s^c_{\dot{a}} +\kb'_{+\,c}\,\kappa'^{+\,a}\,\bar{s}_a^{\dot{a}}\,s^c_{\dot{a}}
+\kb_{+\,a}\,\kappa^{+\,b}\,(\bar{s}')_b^{\dot{a}}\,(s')^a_{\dot{a}} -\kb'_{+\,c}\,\kappa^{+\,c}\,(\bar{s}')_a^{\dot{a}}\,s^a_{\dot{a}}-\kappa_{-\,a}\,\kb'^{-\,a}\,\bar{s}_a^{\dot{b}}\,(s')^b_{\dot{a}}\\ \nn
&& -\kappa_{-\,a}\,\kb^{-\,a}\,(\bar{s}')_b^{\dot{a}}\,(s')^b_{\dot{a}}
+\kappa_{-\,a}\,\kb^{-\,b}\,(\bar{s}')_b^{\dot{a}}\,(s')^a_{\dot{a}}\Big)+\kappa_{-\,a}\,\kb'^{-\,a}\,(\bar{s}')_b^{\dot{a}}\,s^b_{\dot{a}}
-\frac{3}{2}\big(\kappa_{-\,a}\,\kb'^{-\,b}\,(\bar{s}')_b^{\dot{a}}\,s^a_{\dot{a}} \\ \nn
&& +\kb'_{+\,c}\,\kappa^{+\,a}\,(\bar{s}')_a^{\dot{a}}\,s^c_{\dot{a}}\big)+\frac{1}{4}\epsilon^{b\,d}\,\epsilon_{\dot{a}\,\dot{b}}\Big(i\big(
\kappa_{-\,b}\,\kb^{-\,a}\,\bar{s}_a^{\dot{a}}\,(\bar{s}')_d^{\dot{b}}-\kb_{+\,d}\,\kappa^{+\,a}\,\bar{s}_b^{\dot{a}}\,(\bar{s}')_a^{\dot{b}}-3\,\kappa_{-\,a}\,\kb^{-\,a}\,\bar{s}_b^{\dot{a}}\,(\bar{s}')_d^{\dot{b}}\\ \nn
&& +3\,\kb_{+\,a}\,\kappa^{+\,a}\,\bar{s}_b^{\dot{a}}\,(\bar{s}')_d^{\dot{b}}\big)
+2\big(\kappa_{-\,b}\,\kappa'^{+\,a}\,\bar{s}_a^{\dot{a}}\,(\bar{s}')_d^{\dot{b}}-\kb_{+\,d}\,\kb'^{-\,a}\,\bar{s}_b^{\dot{a}}\,(\bar{s}')_a^{\dot{b}}+\kappa'_{-\,d}\,\kappa'^{+\,a}\,\bar{s}_a^{\dot{a}}\,\bar{s}_b^{\dot{b}} \\ \nn
&& -\kappa_{-\,a}\,\kappa'^{+\,a}\,\bar{s}_b^{\dot{a}}\,(\bar{s}')_d^{\dot{b}}
-\kappa_{-\,b}\,\kappa^{+\,a}\,(\bar{s}')_a^{\dot{a}}\,(\bar{s}')_d^{\dot{b}}+\kappa'_{-\,b}\,\kappa^{+\,a}\,\bar{s}_a^{\dot{a}}\,(\bar{s}')_d^{\dot{b}}\big)+6\big(\kb_{+\,b}\,\kb'^{-\,a}\,\bar{s}_a^{\dot{a}}\,(\bar{s}')_d^{\dot{b}}
\\ \nn 
&& -\kb'_{+\,a}\,\kb^{-\,a}\,\bar{s}_b^{\dot{a}}\,(\bar{s}')_d^{\dot{b}}\big)
-8\,\kb'_{+\,d}\,\kb^{-\,a}\,\bar{s}_a^{\dot{a}}\,(\bar{s}')_b^{\dot{b}}\Big)+h.c.
\eea 
Even though quite complicated, both $\mathcal{H}_{BB}$ and $\mathcal{H}_{FF}$
are definitely manageable expressions. Note that the pure bosonic Hamiltonian
suffers from non derivative terms while the pure fermionic do not. For
the latter, these were removed through the shift (\ref{fermshifteta}). For the
bosonic non derivative terms these can be removed through the use of a canonial
transformation as explained in \cite{Frolov:2006cc} and \cite{Sundin:2008vt}. However, for the upcoming analysis,
these will not have any effect on the calculations, so we choose to leave
them as they stand. 

As was explained in the previous section, the exact
form of the fermionic shift relevant for the mixing Hamiltonian has not been
determined. In the appendix we present the original Hamiltonian, prior to
the fermionic shift, together with the form of $\widetilde{\Phi}$. The brave
reader interested in the full mixing Hamiltonian can from there determine the exact form of the additional shift $\hat{\Phi}_3$. Having established
the full shift one can, together with the corrections to the transverse part
of $\PI$, determine the exact form of the shifted $\mathcal{H}_{BF}$.

We have now obtained the relevant Hamiltonian up to quartic order in
number of fields. It is fully gauge fixed and posses the full SU(2$|$2)$\times$U(1)
symmetry of the theory. In the next two sections we will perform explicit calculations
with it, starting by calculating the energy shift for a closed fermionic
subsector and matching these with a set of light-cone Bethe equations.
\section{Fermionic energy shifts and light-cone Bethe equations}
In light of the AdS$_4/$CFT$_3$
correspondence, energies of string excitations should correspond to anomalous
dimensions of single trace operators in certain three dimensional Chern-Simons
theories \cite{Aharony:2008ug}. Based on integrability and the extensive knowledge from the original AdS$_5/$CFT$_4$
correspondence \cite{Beisert:2006ez}, there has been a very rapid progress in understanding how
to encode the spectral problem of both models in terms of Bethe equations. In
\cite{Gromov:2008bz} a all loop set of asymptotic Bethe equations were proposed for the full
OSP($2,2|6)$ model which supposedly encode the energies of all possible
(free) AdS$_4 \times \mathbbm{CP_3}_3$ string configurations. In \cite{Astolfi:2008ji} and \cite{Sundin:2008vt}
it was
shown that the spectrum of string excitations in a closed bosonic subsectors of
the theory exactly match the predictions of the Bethe equations from \cite{Gromov:2008bz}.
In this section we will extend this analysis to include fermionic operators.
Not only will this be an important consistency check of the derived Hamiltonian,
but it will also lend support to the assumed integrability of the full supersymmetric
string model. It is also worth mentioning that this is the first explicit
calculation
probing the higher order fermionic sector of the duality.

Note that we will be rather brief in this section. For readers interested
in the details, we refer to \cite{Hentschel:2007xn} and, especially, \cite{Sundin:2008vt}.
\subsection{Strings in fermionic subsectors}
In this section we will compute the energy shifts for a closed
fermionic subsector constituted of the fields $\kappa^\pm$. Since we have
cubic interaction terms in the Hamiltonian, the standard way to obtain the
energy shifts would be through second order perturbation theory. However, this is quite
an involved procedure since we have to sum over intermediate zeroth order states. A much
simpler approach is to remove the cubic terms through a unitary transformation
of the Hamiltonian \cite{Frolov:2006cc}, \cite{Sundin:2008vt}
\bea 
\label{unitaryexp}
\mathcal{H} \rightarrow e^{i\,V}\,\mathcal{H}\,e^{-i\,V},
\eea
where the guiding principle for the construction of $V$ is that it should
obey
\bea 
\label{propertyV}
i[V,\mathcal{H}_2]=-\mathcal{H}_3,
\eea
and thus removes the unwanted terms.

To find an appropriate generating functional we need the oscillator components of $\mathcal{H}_3$
\bea 
&& \sqrt{g}\,\mathcal{H}_3=\mathcal{H}^{+++}+\mathcal{H}^{++-}+h.c \\ \nn && = \int dk\,dn\,dl\Big(C(k,n,l)^{+++}\,\bar{X}(k)\,\bar{Y}(n)\,\bar{Z}(l)+C(k,n,l)^{++-}\,\bar{X}(k)\,\bar{Y}(n)\,Z(l)\Big)+h.c.
\eea
where the oscillators $X,Y$ and $Z$ takes values in the set of $8_F+8_B$
oscillators. However, since we want the energy shifts for $\kappa^\pm$ excitations,
we only need the piece of $\mathcal{H}_3$ that depends quadratically on $\kappa^\pm$, that is, the first line of (\ref{H3twospinor}). Considering only this part, we can construct
a function $V$ with the property (\ref{propertyV}) as \cite{Frolov:2006cc}
\bea 
&&\sqrt{g}\, V= \int dk\,dn\,dl\Big\{ \\ \nn 
&& \frac{-iC(k,n,l)^{+++}}{w_x(k)+w_y(n)+ w_z(l)}\bar{X}(k)\,\bar{Y}(n)\,\bar{Z}(l)+\frac{-i
C(k,n,l)^{++-}}{w_x(k)+w_y(n)- w_z(l)}\bar{X}(k)\,\bar{Y}(n)\,Z(l)\Big\}+h.c,
\eea
where $w_i(m)$ is either $\omega_m$ or $\Omega_m$ depending on the mass
of $Z(l)$. 
It is straight forward, albeit tedious, to check that this choice
of $V$ indeed removes the cubic terms. However, from (\ref{unitaryexp})
it is clear the $V$ commuted with the cubic part of the Hamiltonian will
give rise to additional quartic terms,
\bea 
\mathcal{H}_4^{Add}=-\frac{1}{2}\{V^2,\mathcal{H}_2\}+V\,\mathcal{H}_2\,V=\frac{i}{2}[V,\mathcal{H}_3].
\eea 
Even though the precise form of $\mathcal{H}_4^{Add}$ is quite
complicated, evaluating its matrix elements is nevertheless significantly
simpler than performing second order perturbation theory with the original
Hamiltonian. Thus, after the unitary transformation, the Hamiltonian is of the form
\bea 
\mathcal{H}=\mathcal{H}_2+\frac{1}{g}\big(\mathcal{H}_4+\mathcal{H}_4^{Add}\big)+\mathcal{O}(g^{-3/2}),
\eea 
and this is the Hamiltonian we will use to calculate energy shifts in first order perturbation theory. 

However, before we move on to that analysis there is one important issue we should comment on - namely, normal ordering. As was the case for the AdS$_5\times$S$^5$ string, the next to leading order piece, which is the cubic contribution in our case, can be assumed to be normal ordered. The subleading piece can, however, not be assumed to be ordered. How to order them is an analysis that we have not performed since to the order of our interest, the normal ordering ambiguities can be addressed using $\zeta$-function regularization, see \cite{Astolfi:2008ji} and \cite{Sundin:2008vt}\footnote{From the point of view of the worldsheet theory, calculating energy shifts to the order we are doing is basically a tree level calculation and the additional effects originating from the ordering terms enter at loop level.}

The states we calculate the energy shifts from will be of the form
\bea 
\label{state}
\ket{m_1\,...\,m_M\,n_1\,...\,n_N}=\bar{c}_1(m_1)\,...\,\bar{c}_1(m_M)\,\bar{d}^2(n_1)\,...\,\bar{d}^2(n_N)\,\ket{0},
\eea
where the sum of the mode numbers has to equal zero, $\sum_{i=1}^M m_i+\sum_{j=1}^N
n_j=0$. For simplicity we only consider states where all mode numbers are
distinct.

The full quartic Hamiltonian, including the additional terms from the unitary
transformation, have a general structure as
\bea 
&& g\,\mathcal{H}_4= \\ \nn 
&& \frac{1}{(2\pi)^2}\int dk\,dn\,dl\,dm\,\delta(m+l-k-n)\Big\{F(k,n,l,m)^{11}_{11}\,\bar{c}_1(k)\,\bar{c}_1(n)\,c^1(l)\,c^1(m)\\
\nn
&& +F(k,n,l,m)^{22}_{22}\,\bar{d}^2(k)\,\bar{d}^2(n)\,d_2(l)\,d_2(m)+F(k,n,l,m)^{12}_{21}\,\bar{c}_1(k)\,\bar{d}^2(n)\,d_2(l)\,c^1(m)\Big\}
\\ \nn 
&& + \textrm{ Non relevant terms.} 
\eea
The components $F(k,n,l,m)_{cd}^{ab}$ are quite complicated functions of
the frequencies and the fermionic wave functions. Luckily, their form gets
constrained considerably when projected on the states (\ref{state}),
\bea 
\label{energyshifts}
&& \Delta E=\bra{n_N\,...\,n_1\,m_M\,...\,m_1}\,\mathcal{H}_4\,\ket{m_1\,...\,m_M\,n_1\,...\,n_N}=
\\ \nn 
&& \frac{1}{g}\Big\{\frac{1}{16}\sum_{i,j=1}^M \frac{\big(m_i-m_j\big)^2}{\omega_{m_i}\,\omega_{m_j}}+\frac{1}{16}\sum_{i, j=1}^N \frac{\big(n_i-n_j\big)^2}{\omega_{n_i}\,\omega_{n_j}}+\frac{1}{8}\sum_{i=1}^M\sum_{j=1}^N\frac{\big(m_i+n_j\big)^2+4\,m_i\,n_j}{\omega_{m_i}\,\omega_{n_j}}\Big\}.
\eea
Since both the $\kappa^\pm$ part of
(\ref{Hff}) and the additional quartic terms are quite complicated, it is a remarkable feature of the uniform light-cone and kappa gauge that
the energy shifts takes such a simple form. 

In the next section we will show that these energy shifts are exactly reproduced
from the asymptotic Bethe equations of \cite{Gromov:2008qe} and \cite{Sundin:2008vt}.
\subsection{Bethe equations}
The starting point of this discussion will be the asymptotic light-cone Bethe
equations of \cite{Sundin:2008vt} given  by\footnote{In \cite{Sundin:2008vt} a large light-cone momentum, $P_+\rightarrow
\infty$ and $\widetilde{\lambda}'\sim$ constant,
expansion was utilized. With the identifications $P_+=2g$ and $\widetilde{\lambda}'=1$, that expansion
is equivalent to the strong coupling expansion used in this paper.}
\bea 
\label{full-lcbe}
&& \Big(\frac{x^+(p_k)}{x^-(p_k)}\Big)^{\frac{1}{2}\big(2g+\eta(M+N)\big)}=
\\ \nn 
&& \Big(\frac{x^+(p_k)}{x^-(p_k)}\Big)^{-g}\prod_{k\neq j}^M\,\Big(\frac{x^+(p_k)-x^-(p_j)}{x^-(p_k)-x^+(p_j)}\Big)^{\frac{1}{2}(1+\eta)}
\sqrt{\frac{1-\big(x^+(p_k)\,x^-(p_j)\big)^{-1}}{1-\big(x^+(p_j)\,x^-(p_k)\big)^{-1}}}
\times \\ \nn 
&& \prod_{j=1}^N\,\Big(\frac{x^+(p_k)-x^-(q_j)}{x^-(p_k)-x^+(q_j)}\Big)^{\frac{1}{2}(1-\eta)}
\sqrt{\frac{1-\big(x^+(q_j)\,x^-(p_k)\big)^{-1}}{1-\big(x^+(p_k)\,x^-(q_j)\big)^{-1}}}+\mathcal{O}(g^{-3}),
\\ \nn 
&& \Big(\frac{x^+(q_k)}{x^-(q_k)}\Big)^{\frac{1}{2}\big(2g+\eta(M+N)\big)}=
\\ \nn 
&& \Big(\frac{x^+(q_k)}{x^-(q_k)}\Big)^{-g}\prod_{k\neq j}^M\,\Big(\frac{x^+(q_k)-x^-(q_j)}{x^-(q_k)-x^+(q_j)}\Big)^{\frac{1}{2}(1+\eta)}
\sqrt{\frac{1-\big(x^+(q_k)\,x^-(q_j)\big)^{-1}}{1-\big(x^+(q_j)\,x^-(q_k)\big)^{-1}}}
\times \\ \nn 
&& \prod_{j=1}^N\,\Big(\frac{x^+(q_k)-x^-(p_j)}{x^-(q_k)-x^+(p_j)}\Big)^{\frac{1}{2}(1-\eta)}
\sqrt{\frac{1-\big(x^+(p_j)\,x^-(q_k)\big)^{-1}}{1-\big(x^+(q_k)\,x^-(p_j)\big)^{-1}}}+\mathcal{O}(g^{-3}).
\eea
\begin{figure}[t]
\centering
\includegraphics[width=0.5\textwidth]{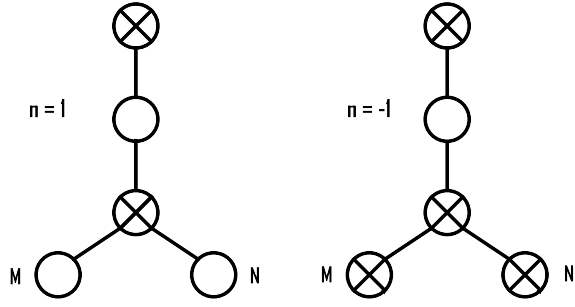}
\caption{Dynkin diagrams for the two choices of gradings, $\eta=\pm 1$}
\label{fig:}
\end{figure}
The constant $\eta=\pm 1$ selects one of the two Dynkin diagrams in figure
\ref{fig:}.
In \cite{Sundin:2008vt} only the Bethe equations for the $\eta=1$ diagram were spelled out but
it should be clear from \cite{Beisert:2005fw} and, especially, \cite{Hentschel:2007xn} how the generalization from
the bosonic case, in \cite{Sundin:2008vt}, to the situation at hand works. The main difference
between the two diagrams is the statistics of the $M$ and $N$ nodes, where
the integers denote the number of $\bar{c}_1$ and $\bar{d}^2$ excitations. For
$\eta=1$ the basic spin flips in the two spin chains are purely bosonic while
for $\eta=-1$ they are fermionic. Since we are calculating energy shifts
for fermionic operators, we need to choose $\eta=-1$\footnote{If not, then
the outer arms of the Bethe equations gets excited and the functional
form of the equations are considerably more complicated. For an example,
see \cite{Hentschel:2007xn}.}.

The spectral parameters $x^\pm(p_k)$ in (\ref{full-lcbe}), can be solved for through
\bea 
\label{spectral}
x^\pm(p_k)+\frac{1}{x^\pm(p_k)}=\frac{1}{h(\lambda)}\Big(\phi(p_k)\pm\frac{i}{2}\Big),
\eea
where
\bea 
\phi(p_k)=\cot\frac{p_k}{2}\sqrt{\frac{1}{4}+4\,h(\lambda)^2\,\sin^2\frac{p_k}{2}}.
\eea
The function $h(\lambda)$ is a novel feature for the AdS$_4$ / CFT$_3$ duality
and is, so far, only known perturbatively\footnote{The reason we could ignore the ordering issues of the light-cone Hamiltonian, is because they kick in at order $\mathcal{O}(\lambda^0)$ of $h(\lambda)$, i.e. beyond the tree level approximation.}
\cite{Nishioka:2008gz}, \cite{Gaiotto:2008cg}, \cite{Grignani:2008is}. It scales differently
in the weak / strong coupling regimes, where in our case we only need the
leading order part of the strong coupling expansion
\bea \label{interpolating}
h(\lambda)=\sqrt{\frac{\lambda}{2}}+\mathcal{O}(\lambda^0),
\eea
where the 't Hooft coupling $\lambda$ is related to $g$ as
\bea 
\lambda=\frac{g^2}{2\pi^2}.
\eea
The two spin chains in (\ref{full-lcbe})
are related through a momentum, or cyclicity, constraint
\bea 
\label{momentumconstraint}
1=\prod_{k=1}^M\frac{x^+(p_j)}{x^-(p_j)}\,\prod_{j=1}^N\frac{x^+(q_j)}{x^-(q_j)}.
\eea
The light-cone energy, corresponding to eigenvalues of $-p_-$ on the string
theory side, can be expressed through
\bea 
\label{lightconeenergyBE}
&& E=\Delta-J= \\ \nn 
&& \sum_{j=1}^M\Big(\sqrt{\frac{1}{4}+4\,h(\lambda)^2\sin^2\frac{p_j}{2}}-\frac{1}{2}\Big)
+\sum_{j=1}^N\Big(\sqrt{\frac{1}{4}+4\,h(\lambda)^2\sin^2\frac{q_j}{2}}-\frac{1}{2}\Big).
\eea
It is the functional form of the light-cone energy that we will match against
the string energy shifts in (\ref{energyshifts}). To achieve this we assume a perturbative expansion for the momentas,
\bea
p_k=\frac{p^0_k}{2g}+\frac{p^1_k}{(2g)^2}, \qquad q_k=\frac{q^0_k}{2g}+\frac{q^1_k}{(2g)^2},
\eea
which allows us to perturbatively solve for $p_k$ and $q_k$ in (\ref{full-lcbe}).
For the leading order contribution one finds
\bea 
p^0_k=4\pi\,m_k, \qquad q_k^0=4\,\pi\,n_k.
\eea
The higher order components $p^1_k$ and $q^1_k$ are a bit more involved but
can straightforwardly be deduced from (\ref{full-lcbe}). 

Using the solutions in (\ref{lightconeenergyBE}) and expanding
gives that the $\mathcal{O}(g^{-1})$ shifts are given by\footnote{We abbreviated
$\omega_{m_k}=\omega_k$ and similar for the $n_k$ indices. Which excitation
the index belong to should be clear from the context.}
\bea 
\label{finalbeshifts}
&&\Delta E = \\ \nn 
&& \frac{1}{4g}\sum_{k=1}^M\Big\{\frac{(M+N)m_k^2}{\omega_k}+\frac{8\,m_k^2}{\omega_k}\Big(\sum_{j=1}^M\frac{m_j(m_k-m_j)}{(1+2\omega_k)(1+2\omega_j)-4\,m_k\,m_j}
\\ \nn 
&&-\sum_{j=1}^N\frac{n_j(m_k-n_j)}{(1+2\omega_k)(1+2\omega_j)-4\,m_k\,n_j}-\sum_{j=1}^N\frac{n_j(1+\omega_j+\omega_k)}{n_j(1+2\omega_k)-m_k(1+2\omega_j)}\Big)\Big\}\\
\nn 
&& +\frac{1}{4g}\sum_{k=1}^N\Big\{\frac{(M+N)n_k^2}{\omega_k}+\frac{8\,n_k^2}{\omega_k}\Big(\sum_{j=1}^N\frac{n_j(n_k-n_j)}{(1+2\omega_k)(1+2\omega_j)-4\,n_k\,n_j}
\\ \nn 
&&-\sum_{j=1}^M\frac{m_j(n_k-m_j)}{(1+2\omega_k)(1+2\omega_j)-4\,n_k\,m_j}-\sum_{j=1}^M\frac{m_j(1+\omega_j+\omega_k)}{m_j(1+2\omega_k)-n_k(1+2\omega_j)}\Big)\Big\},
\eea 
which should be augmented with the expanded cyclicity condition (\ref{momentumconstraint}),
\bea 
\sum_{j=1}^M m_j+\sum_{j=1}^N\,n_j=0.
\eea
After enforcing this constraint in (\ref{finalbeshifts}) one can show
that the energy shifts calculated from the Bethe equations (\ref{full-lcbe})
precisely matches the string energies obtained from diagonalizing the string
Hamiltonian in (\ref{energyshifts}). However, as also was the case for the bosonic calculation in \cite{Sundin:2008vt}, it is quite tedious to show the algebraic equivalence of the two expressions. The use of a computer program able to handle symbolic
manipulations is recommended.  

As a further consistency check of the Hamiltonian we derived, we have also verified the result of \cite{Sundin:2008vt} with the $\eta=1$ grading
using the purely bosonic cubic and quartic Hamiltonian (\ref{H3twospinor})
and (\ref{Hbb}).

This is the first calculation beyond leading order probing the fermionic sector of the AdS$_4 \times \mathbbm{CP_3}_3$ string. Not only does it establish
the validity of the derived Hamiltonian but it also lend strong support to
the assumed integrability, and especially, the factorization properties following from that. 

Before we end this section let us also mention the result in \cite{Dukalski:2009pr}.
There the authors constructed a fermionic reduction to a subsector identical to the SU($1|1)$ sector of the AdS$_5\times$S$^5$ string \cite{Alday:2005jm} \cite{Arutyunov:2005hd}.
However, this is not the sector we have studied since the form of the Bethe equations are not the same as the SU($1|1)\subset$ PSU$(2,2|4)$ light-cone Bethe
equations in \cite{Frolov:2006cc}. The relation between the two sectors is
unclear for us and it would be nice to understand it further. 
\section{Quantum corrections to the heavy modes}
The Bethe equations presented in the earlier section can be extended to the
full symmetry group OSP(2,2$|6$) in which the Bethe roots fall
into short representations of SU(2$|$2), that is, only $4_F+4_B$ modes appear
as fundamental excitation in the scattering matrix. At leading order
these have the magnon dispersion relation, $\omega=\sqrt{\frac{1}{4}+p^2}$,
so it is natural to associate these with the $4_F+4_B$ light string modes, $\kappa^\pm$
and $
\omega_{\dot{a}}$. However, as we have seen, critical string theory exhibits
 $8_F+8_B$ oscillatory degrees
of freedom, so how are we to understand the modes $y, z_i$ and $s^a_{\dot{b}}$?
From the quadratic Lagrangian it certainly seems like they are on an equal
footing as the light modes, so why do they not appear as excitations in the
S-matrix?

By continuing a line of research initiated by Zarembo in \cite{Zarembo:2009au} we
will try to address this question in the upcoming section. We will do this
by calculating loop corrections to the propagators of the massive fields. As
we will argue, the loop corrections have the effect that
the pole gets shifted beyond the energy threshold for pair production of
two light modes, so the heavy state dissolves into a two particle continuum. 

\begin{figure}[t]
\centering
\includegraphics[width=0.9\textwidth]{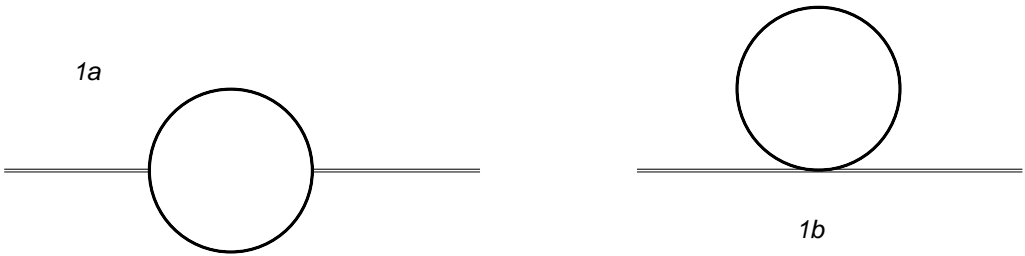}
\caption{Self energy graphs.}
\label{fig:1}
\end{figure}

From the analysis in the previous section, it is clear that the two
type of relevant loop diagrams are a three vertex loop from (\ref{H3twospinor})
and a tadpole diagram from the full quartic Hamiltonian, see Figure \ref{fig:1}. To calculate the
corrections one would need to calculate the full contribution from both types
of diagram. However, for the question wetter the heavy modes come as
fundamental excitations or not, it is enough to focus
our attention on the propagators analytic properties close to the pole.
For the pure quadratic theory, at strictly infinite coupling, the massive propagators has a pole at $\bar{k}^2=1$. Incorporating quantum corrections, it gets shifted as
\bea \nn
\Delta(k)\sim \int d^2k\frac{Z(k)}{\bar{k}^2-1+\frac{1}{g}\delta m+i\epsilon},
\eea
where, as we will show, the relevant part of the mass corrections are of the form
\bea 
\delta m=C(k)\sqrt{1-\frac{1}{\bar{k}^2}}.
\eea
For values of $\bar{k}$
such that the difference $\bar{k}^2-1$ is very small, the first term in the propagator can be as important as the
second one. Since the bare pole lies exactly at the branch point for pair
production of two light modes, the sign of $C(k)$ may change the analytical properties
of $\Delta(k)$. If the sign is positive, then the one particle
pole is shifted below the threshold energy. If negative, however,
the pole gets shifted beyond the threshold energy and disappears. This means
that this field does not exist as a physical excitation for finite values
of the coupling $g$.

As is well known, the behavior of a Feynman integral close to its pole is
dominated by its imaginary part. Thus, the behavior of the quantum corrected
pole can be extracted from the imaginary part
of $\delta m$. This has the pleasant advantage that, for the calculation
at hand, we can neglect the tadpole diagrams. This is easy to understand
if one takes a look at the general structure of such a contribution,
\bea  \nn
\int d^2k \frac{G(k)}{\bar{k}^2-m^2+i\,\epsilon},
\eea 
where $g(k)$ is a even polynomial in $k$ and $m$ is the mass of the particle in
the loop. By direct inspection it is clear that there are no extra branch points
associated to this integral. Of course, there
are however a lot of real terms, both finite and divergent, resulting from
the integral. It is however likely
that supersymmetry guarantees that these terms cancel among themselves\footnote{The
heavy modes are in a semi short representation of the SU$(2|2)$ and should
be BPS protected from mass renormalizations \cite{Zarembo:2009au} \cite{Ahn:2008aa}.}. 

The analysis then boils down to isolating the imaginary part of the three
vertex loops. Since we will only focus on the massive bosonic coordinates,
the relevant part of the cubic Hamiltonian (\ref{H3twospinor}) is
\bea
\label{h3contributing}
\sqrt{g}\,\mathcal{H}_3^{loop}= \big(\bar{\Psi}_a\,\Psi^b\big)'Z^a_b+i\big(\bar{\Psi}\,\gamma^1\,\Psi'
-\bar{\Psi}'\,\gamma^1\,\Psi\big)_a^b\,(Z')^a_b-2i\big(\bar{\Psi}'\,\Psi-\bar{\Psi}\,\Psi'\big)_a^b\,P_{z,b}^a,
\eea
from where its clear that the the fields in the loops are $\omega_{\dot{a}}$
for the singlet and $\kappa^\pm$ for $Z^a_b$.

For the upcoming analysis we will need the bare propagators,
\bea 
&& \bra{0}T\{\Psi(\bar{\sigma})\,\bar{\Psi}(\bar{\sigma}')\}\ket{0}=\frac{i}{2(2\pi)^2}\int d^2p \frac{\big(\gamma^\alpha\,p_\alpha+\frac{1}{2}\big)}{\bar{p}^2-\frac{1}{4}+i\epsilon}e^{-i\bar{p}\cdot\hat{\sigma}},
\\ \nn 
&& \bra{0}T\{\chi(\bar{\sigma})\,\bar{\chi}(\bar{\sigma}')´\}\ket{0}=\frac{i}{2(2\pi)^2}\int d^2p \frac{\big(\gamma^\alpha\,p_\alpha+1\big)}{\bar{p}^2-1+i\epsilon}e^{-i\bar{p}\cdot\hat{\sigma}},
\\ \nn 
&& \bra{0}T\{Z_b^a(\bar{\sigma})\,Z^c_d(\bar{\sigma}')\}\ket{0}=\frac{i}{(2\pi)^2}\int d^2p\frac{\big(2\,\delta^a_d\,\delta^c_b-\delta^a_b\,\delta^c_d\big)}{\bar{p}^2-1+i\epsilon}e^{-i\bar{p}\cdot\hat{\sigma}},
\\ \nn 
&& \bra{0}T\{y(\bar{\sigma})\,y(\bar{\sigma}')\}\ket{0}=\frac{i}{(2\pi)^2}\int d^2p\frac{e^{-i\bar{p}\cdot\hat{\sigma}}}{\bar{p}^2-1+i\epsilon},\\
\nn 
&&
\bra{0}T\{\omega_{\dot{a}}(\bar{\sigma})\,\bar{\omega}^{\dot{b}}(\bar{\sigma}')\}\ket{0}=\frac{2\,i}{(2\pi)^2}\int d^2p\frac{\delta_{\dot{a}}^{\dot{b}}}{\bar{p}^2-\frac{1}{4}+i\epsilon} e^{-i\bar{p}\cdot\hat{\sigma}},
\eea 
where we abbreviated $\hat{\sigma}=\bar{\sigma}-\bar{\sigma}'$ and $\bar{\sigma}=(\tau,\sigma)$.
\subsection{Massive singlet}
We will start the analysis with the massive singlet $y$, already calculated
by Zarembo in \cite{Zarembo:2009au}. The analysis basically boils down to determining the sign
of the mass correction and since we will encounter (complex) multi valued functions, some care is asked for when determining which value to take as physical. For this reason we will be rather detailed in this part of the
calculation. 

For the singlet we find that one loop corrected
propagator equals
\bea 
&& \bra{\Omega}T\big(y(x)\,y(y)\big)\ket{\Omega}=\frac{i}{(2\pi)^2}\int d^2k\frac{e^{-i\bar{k}\cdot(\bar{x}-\bar{y})}}{\bar{k}^2-1+i\epsilon}\Big(1-\frac{1}{\bar{k}^2-1-i\epsilon}\PI_{00}\Big),
\eea
where the polarization tensor is given by
\bea 
-\PI_{00}=\frac{i}{2(2\pi)^2}\int d^2p\frac{\big(2p_0-k_0\big)^2}{\big(\bar{p}^2-\frac{1}{4}+i\epsilon\big)\big((\bar{p}-\bar{k})^2-\frac{1}{4}+i\epsilon\big)}
\eea 
Using the standard Feynman parameterization with $\bar{q}=\bar{p}-\bar{k}\,z$,
a direct computation gives 
\bea
\label{pi00}
&&-\PI_{00}= \\ \nn 
&&\frac{1}{4\pi}\Big\{\frac{1}{\eta}-\gamma-\log(\pi)-\int_0^1 dz\Big(\log\big(\frac{1}{4}-\bar{k}^2(1-z)z+i\epsilon\big)
+\frac{(1-2z)^2k_0^2}{2(\frac{1}{4}-\bar{k}^2(1-z)z+i\epsilon)}\Big)\Big\},
\eea
where we used dimensional regularization to isolate the divergence. For a purely real argument the logarithm develops a imaginary
$\pm i \pi$ part when $\bar{k}^2>1$, and to isolate it, we integrate $z$
over the interval $\frac{1}{2}(1\pm\sqrt{1-\frac{1}{\bar{k}^2}})$. With the $\epsilon$ prescription included, we find that it gives rise to
a small positive imaginary contribution, so it is the $i \pi$ part of Im
$(\log)$ that we should use. Thus, for $\bar{k}^2> 1$, its imaginary contribution
is
\bea 
\textrm{Im}\Big[\frac{1}{4\pi}\int \,\log(\frac{1}{4}-\bar{k}^2(1-z)z)\Big]=\frac{1}{4}\,\sqrt{1-\frac{1}{\bar{k}^2}}.
\eea
If we introduce the short hand notation $\alpha=\frac{1}{2}\sqrt{1-\frac{1}{\bar{k}^2}-\frac{4\,i\epsilon}{\bar{k}^2}}$
and shift $z\rightarrow y+\frac{1}{2}$,
the last term in (\ref{pi00}) can be written as
\bea 
\label{integral1}
\frac{k_0^2}{4\pi\,\bar{k}^2}\int_0^{\frac{1}{2}}dy\big(4+\frac{2\alpha}{y-\alpha}-\frac{2\alpha}{y+\alpha}\big).
\eea
The imaginary part of this integral comes from the middle term, where the
$\epsilon$ prescription gives a negative imaginary contribution. To calculate the imaginary part of (\ref{integral1}) we introduce $y-\alpha=\epsilon_0\,e^{i\theta}$,
which gives
\bea
-\frac{k_0^2}{4\,\bar{k}^2}\sqrt{1-\frac{1}{\bar{k}^2}}=-\frac{k_0^2}{4}\sqrt{1-\frac{1}{\bar{k}^2}}+\mathcal{O}(1-\frac{1}{\bar{k}^2})^{\frac{3}{2}},
\eea
where we assumed that $\bar{k}^2$ is close to the two particle threshold.

Combining the two results shows that
\bea 
\textrm{Im}\,\PI_{00}=\frac{1}{4}(1-k_0^2)\sqrt{1-\frac{1}{\bar{k}^2}}=-\frac{1}{4}k_1^2\,\sqrt{1-\frac{1}{\bar{k}^2}}+\mathcal{O}(1-\frac{1}{\bar{k}^2})^{\frac{3}{2}},
\eea
which is negative definite close to the pole. This is almost what Zarembo calculated in \cite{Zarembo:2009au}. The difference lies in the form of the square root, which in \cite{Zarembo:2009au} was, $\sqrt{1-\bar{k}^2}$, while we have $\sqrt{1-\frac{1}{\bar{k}^2}}$.
This is related to the expansion scheme and has no physical consequence.
 What
is important is the presence of a positive definite function with the correct
overall sign in front. 
\subsection{Massive AdS coordinates}
Having established what happens to the singlet when loop corrections are taken
into account we turn next to the remaining massive coordinates. The corrected
propagator we want to calculate is
\bea
\bra{\Omega}T\big(Z_l^k(x)\,Z^m_n(y)\big)\ket{\Omega}.
\eea
For this calculation, it is convenient to write the relevant part of the cubic Hamiltonian, (\ref{H3twospinor}), as
\bea 
\sqrt{g}\,\mathcal{H}_3=i\big(\bar{\Psi}\,\gamma^1\,\Psi'-\bar{\Psi}'\gamma^1\,\Psi+i\,\bar{\Psi}\cdot\Psi\big)^b_a
(Z')^a_b-2i\big(\bar{\Psi}'\,\Psi-\bar{\Psi}\,\Psi'\big)^b_a(P_z)_b^a.
\eea
Due to the fermions in the loop, we will encounter quadratic divergences along
the way. However, as was the case for the singlet, these will not contribute
to the imaginary part. 

Due to the more complicated cubic Hamiltonian, the
calculation will be more involved. However, pushing through with the calculation and using the Feynman parameterization as before,
gives that the relevant terms are of the form
\bea 
\delta m=\int\,d^2q\frac{F_0(\bar{k})+F_2(\bar{k},q_0^2,q_1^2)+F_4(\bar{k},q_0^2\,q_1^2,q_1^4)}{(\bar{q}^2-\bar{k}^2(1-z)z-\frac{1}{4}-i\,\epsilon)^2},
\eea
where the subscript denote the power of $q_i$ in the nominator.

To determine the form of the functions $F_i$, we repeat the same procedure
as for the singlet computation. Unfortunately they are rather involved so we will not present them explicitly, but a straight
forward, albeit somewhat tedious, calculation shows that
\bea 
&& \delta m_0=-2\,k_1^2\sqrt{1-\frac{1}{\bar{k}^2}}, \qquad \delta m_2=\frac{1}{3}\big(k_0^2-k_0^4+4k_1^2+k_1^4)\sqrt{1-\frac{1}{\bar{k}^2}},\\
\nn 
&& \delta m_4=\frac{1}{3}(\bar{k}^2-1)(k_0^2+k_1^2)\sqrt{1-\frac{1}{\bar{k}^2}},
\eea
which added together gives 
\bea 
\delta m =-k_1^2\sqrt{1-\frac{1}{\bar{k}^2}}\big(2\delta^k_n\,\delta^m_l-\delta^k_l\,\delta^m_n\big),
\eea
which is strictly negative\footnote{Or, to be precise, it is strictly negative
when we restrict to the $z_i$ propagation, $<\,T(z_i\,z_j)\,>$.} and exact for $\bar{k}^2>1$. 

With this we conclude that all the massive bosons dissolve in a two particle continuum.
\subsection{Massive fermions and comments}
Even though we have not performed the calculation in detail, it is plausible
that the massive $s_{\dot{a}}^b$ fields exhibit the same property as the
massive bosons. By direct inspection of the cubic Hamiltonian it is clear
that the fields in the loop will be the two light $\omega_{\dot{a}}$ and
$\kappa^\pm$. Unfortunately, due to the rather entangled mixing between the
$s^b_{\dot{a}}$ and $\kappa^\pm$ fields, the imaginary part of the propagator
is rather involved. Nevertheless, it is still of the form
\bea 
\nn 
C(k)\sqrt{1-\frac{1}{\bar{k}^2}}+\mathcal{O}(1-\frac{1}{\bar{k}^2})^{\frac{3}{2}},
\eea
with a complicated $C(k)$ which we have not determined. Instead of pursuing this line of research, a much better way to approach the problem would be to
calculate the worldsheet scattering matrix and from there study the behavior of the
massive fields. Unfortunately, since it is only through loop corrections
that the physical role of the massive fields emerge, the calculation of the
scattering matrix would be complicated. In fact, not even for the AdS$_5 \times$S$^5$
case is the one loop BMN scattering matrix fully known. This gives a rather
grim outlook for the possibility of deriving the exact one loop behavior
of the AdS$_4\times\mathbbm{CP_3}_3$ BMN string. 
\section{Summary and closing comments}
In this paper we have presented a detailed discussion about the type IIA
superstring in AdS$_4\times\mathbbm{CP_3}_3$. By starting directly from the
$\mathfrak{osp}(2,2|6)$ superalgebra we constructed the string Lagrangian through its graded components. Then by identifying the bosonic subalgebra commuting with the light-cone
Hamiltonian, we introduced a notation covariant under the gauge fixed SU$(2|2)\times$
U(1) symmetry. The covariant string Lagrangian was the starting point for
a perturbative analysis in a strong coupling limit where we almost immediately 
ran into problem due to the presence of higher order kinetic terms for the fermionis.
These had the sad effect that they complicated the general structure of the
theory to such an extent that we only presented parts of the canonical Hamiltonian.
Nevertheless, we
proceeded with a calculation of energy shifts for fermionic string configurations
built out of a arbitrary number of $c_1$ and $d^2$ oscillators. These shifts
we successfully matched with the prediction coming from a conjectured
set of light-cone Bethe equations. 

We then moved on to an investigation of the role of the massive bosonic modes.
By calculating loop corrections to the propagators of the massive fields
we saw that the massive modes dissolved into a two particle continuum. 

We also provided an extensive
appendix where the original Hamiltonian, including the kinetic terms of the
fermions were spelled out in detail. 

All in all we have presented a rather thorough study of the AdS$_4\times\mathbbm{CP_3}_3$
superstring. Naturally a lot remains to be done, where perhaps the most stressing,
at least from the point of view of our analysis, is to establish the one
loop scattering matrix for the heavy modes. Even though we provided arguments
for that the heavy modes dissolve in a two particle continuum, it would be
desirable to see it explicitly in terms of Feynman diagrams. Unfortunately,
due to the complexity of the theory, it does not seem very plausible that one can achieve this through the use of
the BMN string. Perhaps a better way to approach the problem would be through
the so called near flat space limit \cite{Maldacena:2006rv}, \cite{Kreuzer:2008vd}. 

Another interesting line of research would be to consider higher order corrections
to the interpolating function $h(\lambda)$ that occurs in the magnon dispersion relation.
It has been extensively studied in \cite{Alday:2008ut} \cite{McLoughlin:2008ms} \cite{Krishnan:2008zs} \cite{McLoughlin:2008he}, but its higher order structure remains
unknown. 

We plan to return to some of these questions in further investigations.
\acknowledgments
I would like to begin with thanking Dmitri Bykov for a early collaboration on this project and many subsequent illuminating discussions. I would also like to thank the following persons for useful discussions, S. Frolov, C. Kalousios, T. McLoughlin, J.Plefka, V.G.M Puletti, A. Rodigast, R. Suzuki, D. Young and K. Zarembo. I would also like to thank the School of Mathematics at Trinity College for hospitality where parts of this work were carried out. Finally I would like to thank the organizers of the conference Integrability in Gauge and String Theory for providing a stimulating research environment. This work was in part supported by a grant from the International Max-Planck Research School for Geometric Analysis, Gravitation and String Theory.
\newpage
\begin{appendix}
\section{Matrices}
In this appendix we present the exact form of some of the matrices encountered
in the main text. The conventions are those of Arutyunov and Frolov in \cite{Arutyunov:2008if}.

Starting out with the SO(1,3) $\Gamma$ we have
\begin{displaymath}
\Gamma^0=\left( \begin{array}{cccc}
1  & 0 & 0 & 0 \\
0 & 1 & 0 & 0 \\
0  & 0 & -1 & 0 \\
0  & 0 & 0& -1 \\
\end{array} \right), 
\Gamma^1=\left( \begin{array}{cccc}
0  & 0 & 0 & 1 \\
0 & 0 & 1 & 0 \\
0  & -1 & 0 & 0 \\
-1  & 0 & 0& 0  \\
\end{array} \right), \\
\Gamma^2=\left( \begin{array}{cccc}
0  & 0 & 0 & -i \\
0 & 0 & i & 0 \\
0  & i & 0 & 0 \\
-i  & 0 & 0& 0 \\
\end{array} \right),
\Gamma^3=\left( \begin{array}{cccc}
0  & 0 & 1 & 0 \\
0 & 0 & 0 & -1 \\
-1  & 0 & 0 & 0 \\
0  & 1 & 0& 0 \\
\end{array} \right),
\end{displaymath}
satisfying $\{\Gamma^\mu,\Gamma^\nu\}=2\eta^{\mu \nu}$ with signature (+,-,-,-).
As usual, the combinations $\Gamma^{\mu\nu}=\frac{1}{2}[\Gamma^\mu,\Gamma^\nu]$ generate
the lie algebra $\mathfrak{so}(1,3)$.

The charge conjugation matrix, $C_4$, can be expressed in terms of the $\Gamma$
matrices as
\begin{displaymath}
C_4=i\,\Gamma^0\,\Gamma^2=\left( \begin{array}{cccc}
0  & 0 & 0 & 1 \\
0 & 0 & -1 & 0 \\
0  & 1 & 0 & 0 \\
-1  & 0 & 0& 0 \\
\end{array} \right).
\end{displaymath}
The six $T_i$ matrices are generators of $\mathfrak{so}(6)$ along $\mathbbm{CP_3}_3$
and are given by
\bea
&& T_1=E_{13}-E_{31}-E_{24}+E_{42}, \qquad T_2=E_{14}-E_{41}+E_{23}-E_{32},\\
\nn 
&& T_3=E_{15}-E_{51}-E_{26}+E_{62}, \qquad T_2=E_{16}-E_{61}+E_{25}-E_{52},
\\ \nn 
&& T_5=E_{35}-E_{53}-E_{46}+E_{64}, \qquad T_2=E_{36}-E_{63}+E_{45}-E_{54},
\eea
where $E_{ij}$ is the $6 \times 6$ matrix with all elements zero except the
$i,j$'th component which is unity. The normalization is as follows,
\bea 
Tr (T_i\,T_j)=-4\,\delta_{ij}.
\eea
The $T_i$ matrices satisfy the following important properties,
\bea 
\{T_1,T_2\}=0, \qquad \{T_3,T_4\}=0, \qquad \{T_5,T_6\}=0.
\eea
In the text we frequently make use of the complex combinations,
\bea 
\tau_1=\frac{1}{2}\big(T_1-i\,T_2\big),\qquad \tau_2=\frac{1}{2}\big(T_3-i\,T_4\big),
\eea 
and $\bar{\tau}_i$ for conjugated combinations.
\section{Mixing term of the original Hamiltonian}
In this appendix we present the full, non shifted, quartic Hamiltonian, which
combined
 with the fermionic kinetic term in (\ref{kineticfermL}) encodes the
 full dynamics of the quartic theory. 

We start out by presenting the original cubic Hamiltonian which is similar but
not identical to the shifted one,
\bea
\label{orgcubic}
&& \sqrt{g}\,\mathcal{H}_3^{ns}=\\ \nn 
&&\big(\bar{\Psi}'\cdot \Psi\big)^a_b\,Z^b_a +\big(\bar{\Psi}\,\gamma^1\,\Psi'\big)^a_b\,(Z')^b_a
+i\,y\,\omega_{\dot{a}}\,\bar{p}^{\dot{a}} +3i\,(\kappa_{-\,a}\,\bar{s}^{a\,\dot{a}}-\kb_{+\,a}\,s^{a\,\dot{a}}\big)\,p_ {\dot{a}}\\ \nn 
&& +\big(\frac{3}{8}(\kb_{+\,a}\,s^{a\,\dot{a}}+\kappa_{-\,a}\,\bar{s}^{a\,\dot{a}})+i (\kappa'_ {-\,a}\,s^{a\,\dot{a}}-\kb'_ {+\,a}\,\bar{s}^{a\,\dot{a}})+\frac{i}{2}(\kb_{+\,a}\,(\bar{s}')^{a\,\dot{a}}-\kappa_{-\,a}\,(s')^{a\,\dot{a}})\big)\,\omega_{\dot{a}}\\ \nn
&&
+\frac{1}{2}\big(\kappa_{-\,a}\,(\bar{s}')^{a\,\dot{a}}-\kappa'_ {-\,a}\,\bar{s}^{a\,\dot{a}}-\kb'_ {+\,a}\,s^{a\,\dot{a}}+\kb_{+\,a}\,(s')^{a\,\dot{a}}\big)\,\omega'_ {\dot{a}}+h.c,
\eea
where the $ns$ superscript denotes that this is the non shifted Hamiltonian.

Next we turn to the quartic interactions, where we as before split up the Hamiltonian
according to its field content. The pure bosonic part will naturally be identical
to (\ref{Hbb}) so we will not present it again. For the pure fermionic part
we find
\bea 
&& g\,\mathcal{H}_{FF}^{ns}=\kappa_{-\,a}\,\kb_{+\,b}\,\kappa^{+\,b}\,\kb^{-\,a}-\kappa_{-\,a}\,\kb_{+\,b}\,\kappa'^{+\,a}\,\kb'^{-\,b}-\kappa_{-\,a}\,\kappa'_{-\,b}\,\kappa^{+\,b}\,\kappa'^{+\,a}\\
\nn 
&& -\kappa_{-\,a}\,\kappa'_{-\,b}\,\kb^{-\,b}\,\kb'^{-\,a}-\kappa_{-\,a}\,\kb'_{+\,b}\,\kappa^{+\,b}\,\kb'^{-\,a}
-\kappa_{-\,a}\,\kb'_{+\,b}\,\kb^{-\,b}\,\kappa'^{+\,a}-\kb_{+\,a}\,\kappa'_{-\,b}\,\kappa^{+\,b}\,\kb'^{-\,a}\\
\nn
&& -\kb_{+\,a}\,\kb'_{+\,b}\,\kappa^{+\,b}\,\kappa'^{+\,a}+\frac{1}{2}\big(\kappa_{-\,a}\,\kappa_{-\,b}\,\kb^{-\,a}\,\kb^{-\,b}-\kappa_{-\,a}\,\kappa_{-\,b}\,\kb'^{-\,a}\,\kb'^{-\,b}
+\kb_{+\,a}\,\kb_{+\,b}\,\kappa^{+\,a}\,\kappa^{+\,b} \\ \nn
&& -\kb_{+\,a}\,\kb_{+\,b}\,\kappa'^{+\,a}\,\kappa'^{+\,b}\big)-\frac{3}{2}\big(\kappa_{-\,a}\,\kappa_{-\,b}\,\kappa'^{+\,a}\,\kappa'^{+\,b}+\kb_{+\,a}\,\kb_{+\,b}\,\kb'^{-\,a}\,\kb'^{-\,b}\big)
+2\big(\kappa_{-\,a}\,\kappa'_{-\,b}\,\kappa^{+\,a}\,\kappa'^{+\,b}\\ \nn
&& -\kappa_{-\,a}\,\kb_{+\,b}\,\kappa^{+\,a}\,\kb^{-\,b}+\kappa_{-\,a}\,\kb'_{+\,b}\,\kb^{-\,a}\,\kappa'^{+\,b}+\kb_{+\,a}\,\kappa'_{-\,b}\,\kappa^{+\,a}\,\kb'^{-\,b}\big)
+3\,\kappa_{-\,a}\,\kb_{+\,b}\,\kappa'^{+\,b}\,\kb'^{-\,a} \\ \nn 
&& +i\Big(\kappa_{-\,a}\,\kb_{+\,b}\,\kb^{-\,b}\,\kb'^{-\,a}+\kappa_{-\,a}\,\kb_{+\,b}\,\kappa^{+\,a}\,\kappa'^{+\,b}
+\frac{1}{2}\big(\kappa_{-\,a}\,\kb_{+\,b}\,\kappa^{+\,b}\,\kappa'^{+\,a}+\kappa_{-\,a}\,\kb_{+\,b}\,\kb^{-\,a}\,\kb'^{-\,b}\big)
\\ \nn 
&& -3i\big(\kappa_{-\,a}\,\kappa_{-\,b}\,\kappa^{+\,a}\,\kb'^{-\,b}+\kappa_{-\,a}\,\kb'_{+\,b}\,\kappa^{+\,a}\,\kappa^{+\,b}\big)
+i\frac{7}{2}\big(\kb_{+\,a}\,\kb_{+\,b}\,\kappa^{+\,a}\,\kb'^{-\,b}-\kappa_{-\,a}\,\kappa_{-\,b}\,\kb^{-\,a}\,\kappa'^{+\,b}\big)\Big)\\
\nn 
&& +\frac{i}{2}\epsilon^{\dot{a}\,\dot{b}}\,\epsilon_{a\,b}\big(s^{a}_{\dot{a}}\,s^{c}_{\dot{b}}\,s^{b}_{\dot{d}}\,(\bar{s}')^{\dot{d}}_c
+\frac{1}{2}(s^c_{\dot{a}}\,s^a_{\dot{b}}\,(s')^b_{\dot{d}}\,\bar{s}_c^{\dot{d}}-s^a_{\dot{d}}\,s^b_{\dot{a}}\,(s')^c_{\dot{b}}\,\bar{s}_c^{\dot{d}})\big)+\epsilon^{\dot{a}\,\dot{b}}\,\epsilon_{b\,c}\big(-i\,s^a_{\dot{a}}\,s^b_{\dot{b}}(\kappa_{-\,a}\,\kb'^{-\,c}\\
\nn 
&& +\kb'_{+\,a}\,\kappa^{+\,c}-\frac{i}{2}\,\kb'_{+\,a}\,\kb'^{-\,c})+\frac{1}{2}s^b_{\dot{a}}\,(s')^a_{\dot{b}}(\kb_{+\,a}\,\kb'^{-\,c}+\kb'_{+\,a}\,\kb^{-\,c}-i\,\kappa_{-\,a}\,\kb^{-\,c}-i\,\kb_{+\,a}\,\kappa^{+\,c})\\
\nn 
&& +\frac{1}{2}s^b_{\dot{a}}\,(s')^c_{\dot{b}}\big(i\,\kb_{+\,a}\,\kappa^{+\,a}-\kb'_{+\,a}\,\kb^{-\,a}\big)-\frac{1}{2}(s')^a_{\dot{a}}\,(s')^b_{\dot{b}}\,\kb_{+\,a}\,\kb^{-\,c}\big)
+s^a_{\dot{a}}\,\bar{s}_a^{\dot{a}}\big(i\,\kb_{+\,b}\,\kb'^{-\,b}-i\,\kappa'_{-\,b}\,\kappa^{+\,b}\\
\nn 
&& +\kb_{+\,b}\,\kappa^{+\,b}-\frac{1}{2}\kappa'_{-\,b}\,\kb'^{-\,b}-\frac{5}{4}\kappa_{-\,b}\,\kb^{-\,b}\big)+s^a_{\dot{a}}\,\bar{s}_b^{\dot{a}}\big(i\,\kappa_{-\,a}\,\kappa'^{+\,b}-i\,\kb_{+\,a}\,\kb'^{-\,b}
+\frac{1}{2}\kappa'_{-\,a}\,\kb'^{-\,b}+\frac{1}{2}\kb'_{+\,a}\,\kappa'^{+\,b}\\
\nn 
&&+\frac{1}{4}\kappa_{-\,a}\,\kb^{-\,b}+\frac{1}{4}\kb_{+\,a}\,\kappa^{+\,b}\big)+(s')^a_{\dot{a}}\,\bar{s}_a^{\dot{a}}\big(
\kappa'_{-\,b}\,\kb^{-\,b}+\frac{i}{2}\kb_{+\,b}\,\kb^{-\,b}-\frac{1}{2}\kappa_{-\,b}\,\kb'^{-\,b}-\frac{1}{2}\kb_{+\,b}\,\kappa'^{+\,b}+\frac{1}{2}\kb'_{+\,b}\,\kappa^{+\,b}\big)\\
\nn 
&& -(s')^a_{\dot{a}}\,\bar{s}^{\dot{a}}_b\big(i\,\kb_{+\,a}\,\kb^{-\,b}+\frac{1}{2}\kb_{+\,a}\,\kappa'^{+\,b}+\frac{1}{2}\kappa'_{-\,a}\,\kb^{-\,b}\big)-\frac{1}{2}(s')^a_{\dot{a}}\,(\bar{s}')_a^{\dot{a}}\,\kappa_{-\,b}\,\kb^{-\,b}\\
\nn 
&& +\frac{1}{2}(s')^a_{\dot{a}}\,(\bar{s}')^{\dot{a}}_b\big(\kappa_{-\,a}\,\kb^{-\,b}+\kb_{+\,a}\,\kappa^{+\,b}\big)+h.c.
\eea
The original mixing Hamiltonian is rather involved and is given by
\bea \label{quarticNS}
&& -g\,\mathcal{H}^{ns}_{BF}=\\ \nn 
&&
\frac{i}{2}y^2\,s^a_{\dot{a}}\,(s')^{\dot{a}}_a-y\,s^a_{\dot{a}}\,(\bar{s}')^{\dot{a}}_b\,Z^b_a
-\frac{i}{4}y^2\,\bar{\Psi}\,\gamma^1\,\Psi'
- p_y^2\big(2\,\bar{\Psi}\cdot\Psi+i\,\bar{\Psi}'\,\gamma^1\,\Psi\big)
-\frac{1}{2}y'^2\big(\bar{\Psi}\cdot\Psi
 +\frac{i}{2}\bar{\Psi}'\,\gamma^1\,\Psi\big) \\ \nn 
 && -i\frac{3}{4}\,p_y\,\omega_{\dot{a}}\,\kappa_{-\,a}\,\bar{s}^{a\,\dot{a}}+3\big(\frac{i}{2}\,y\,p_{\dot{a}}\,\kappa_{-\,a}\,\bar{s}^{a\,\dot{a}}
 -\frac{1}{16}\,y\,\omega_{\dot{a}}\,\kappa_{-\,a}\,\bar{s}^{a\,\dot{a}}\big)+\frac{i}{4}\,y\,\omega_{\dot{a}}\,\kb_{+\,a}\,(\bar{s}')^{a\,\dot{a}}
 -\frac{i}{2}\,y\,\omega_{\dot{a}}\,\kb'_{+\,a}\,\bar{s}^{a\,\dot{a}}\\ \nn
 &&+\frac{i}{4}\,y\,\omega'_{\dot{a}}\,\kappa_{-\,a}\,(\bar{s}')^{a\,\dot{a}}-\frac{i}{4}\,y\,\omega'_{\dot{a}}\,\kappa'_{-\,a}\,\bar{s}^{a\,\dot{a}}
 +\frac{i}{4}\,y'\,\omega_{\dot{a}}\,\kappa'_{-,a}\,\bar{s}^{a\,\dot{a}}-\frac{i}{4}\,y'\,\omega_{\dot{a}}\,\kappa_{-\,a}\,(\bar{s}')^{a\,\dot{a}}
 -i\,\frac{3}{4}\,p_y\,\omega_{\dot{a}}\,\kb_{+\,a}\,s^{a\,\dot{a}} \\ \nn
 && +i\,\frac{3}{2}\,y\,p_{\dot{a}}\,\kb_{+\,a}\,s^{a\,\dot{a}}+i\,\frac{3}{16}\,y\,\omega_{\dot{a}}\,\kb_{+\,a}s^{a\,\dot{a}}
 -\frac{i}{2}\,y\,\omega_{\dot{a}}\,\kappa'_{-\,a}\,s^{a\,\dot{a}}+\frac{i}{4}\,y\,\omega_{\dot{a}}\,\kappa_{-\,a}\,(s')^{a\,\dot{a}}
 +\frac{1}{4}\,y\,\omega'_{\dot{a}}\,\kb'_{+\,a}\,s^{a\,\dot{a}}\\ \nn 
 &&-\frac{1}{4}\,y\,\omega'_{\dot{a}}\,\kb_{+\,a}\,(s')^{a\,\dot{a}}-\frac{1}{4}\,y'\,\omega_{\dot{a}}\,\kb'_{+\,a}\,s^{a\,\dot{a}}
 +\frac{1}{4}\,y'\,\omega_{\dot{a}}\,\kb_{+\,a}\,(s')^{a\,\dot{a}}+\omega_{\dot{a}}\big(\kb'_{+\,a}\,s^{b\,\dot{a}}+\frac{1}{2}\kb_{+\,a}\,(s')^{b\,\dot{a}}\big)\,Z^a_b\\ \nn
 && -2\,i\,\bar{\Psi}_a\,\gamma^0\,\Psi^b\,\big(P_z\cdot
 Z-\frac{1}{2}Tr(P_z\cdot Z)\mathbbm{1}\big)_b^a -\bar{\Psi}_a\cdot (\Psi')^b\,\big(Z\cdot Z'-\frac{1}{2}Tr(Z\cdot Z')\mathbbm{1}\big)_b^a \\ \nn 
&&
 -\frac{1}{2}\big(i\,\bar{\Psi}'\,\gamma^1\,\Psi +2\,\bar{\Psi}\cdot\Psi\big)\,Tr(Pz\cdot
 Pz)-\frac{i}{8}\,\bar{\Psi}'\,\gamma^1\,\Psi\,Tr(Z\cdot
 Z)-\frac{i}{4}\,s^{a}_{\dot{a}}\,(s')_{a}^{\dot{a}}\,Tr(Z\cdot
 Z)\\ \nn 
 &&-\frac{1}{8}\big(i\,\bar{\Psi}'\,\gamma^1\,\Psi+2\,\bar{\Psi}\cdot\Psi\big)\, Tr(Z'\cdot Z')+\frac{1}{2}(s')^{a}_{\dot{a}}\,\bar{s}_b^{\dot{a}}\big(Z\cdot Z'-\frac{1}{2}Tr(Z\cdot
 Z')\mathbbm{1}\big)^b_a \\ \nn 
&&  +2i\,s^{a}_{\dot{a}}\,\bar{s}_b^{\dot{a}}\big(P_z\cdot Z-\frac{1}{2}Tr(P_z\cdot
 Z)\mathbbm{1}\big)^b_a-\big(\kappa'_{-\,a}\,\bar{s}_b^{\dot{a}}+\frac{1}{2}\kappa_{-\,a}\,(\bar{s}')^{\dot{a}}_b\,\big)\omega_{\dot{a}}\,Z^{b\,a}
 +2i\,x'^-\,\bar{\Psi}\,\gamma^0\,\Psi'\\ \nn 
 && -4\big(i\,\bar{\Psi}'\,\gamma^1\,\Psi+2\,\bar{\Psi}\cdot \Psi\big)\,p_{\dot{a}}\,\bar{p}^{\dot{a}} +2i\,\bar{\Psi}\,\gamma^0\,\Psi\,p_{\dot{a}}\,\bar{\omega}^{\dot{a}} -\frac{1}{4}\big(i\,\bar{\Psi}'\,\gamma^1\,\Psi+2\,\bar{\Psi}\cdot\Psi\big)\,\omega'_{\dot{a}}\,(\bar{\omega}')^{\dot{a}}
 \\ \nn 
 && -i\,\frac{9}{16}\,\bar{\Psi}\,\gamma^1\,\Psi'\,\omega_{\dot{a}}\,\bar{\omega}^{\dot{a}}
 +\frac{1}{2}\big(\bar{\Psi}'\cdot\Psi-\bar{\Psi}\cdot\,\Psi'\big)\omega'_{\dot{a}}\,\bar{\omega}^{\dot{a}}+\frac{1}{4}\big(s^a_{\dot{a}}\,(\bar{s}')_a^{\dot{b}}-(s')^a_{\dot{a}}\,\bar{s}_a^{\dot{b}}\big)\,\omega_{\dot{b}}\,(\bar{\omega}')^{\dot{a}}
 \\ \nn 
&&+\frac{1}{8}\big(s^a_{\dot{a}}\,(\bar{s}')_a^{\dot{a}}-s^a_{\dot{a}}\,\bar{s}_a^{\dot{a}}\big)\,\omega'_{\dot{b}}\,\bar{\omega}^{\dot{b}} +2i\,s^a_{\dot{a}}\,s^{\dot{b}}_a\,\omega_{\dot{b}}\,\bar{p}^{\dot{a}}
-i\,s^a_{\dot{a}}\,\bar{s}_a^{\dot{a}}\,\omega_{\dot{b}}\,\bar{p}^{\dot{b}}+\frac{i}{4}\,s^a_{\dot{a}}\,(s')^{\dot{a}}_a\,\omega_{\dot{b}}\,\bar{\omega}^{\dot{b}}+h.c.
\eea 
Note the slight asymmetry between the $\kappa^\pm$ fields. This is due to
the fact that we have not considered the kinetic terms of the fermions, with witch one should augment the non shifted Hamiltonian.
\section{Fermionic shift}
The fermionic shift has to be implemented on the quadratic and cubic Hamiltonian
in
(\ref{quadraticL}) and (\ref{orgcubic}). In order to attain this one need
the explicit form of the fermionic shift. Starting from (\ref{kineticfermL}),
one can write
\bea 
\label{formoffermshift}
&& \frac{1}{g}\mathscr{L}^\eta_{Kin}=\frac{1}{2}Str\,\dot{\eta}\Big\{[\eta,G_t\,\PI\,G_t^{-1}]+\frac{1}{4}\big([G_t\,\PI\,G_t^{-1},\eta^3]
+\eta\,[G_t\,\PI\,G_t^{-1},\eta]\,\eta\big) \\ \nn 
&&-i\,\kappa\,G_t\,\Upsilon\,G_t^{-1}\Big(\frac{i}{2}[\Sigma_-,\eta]x'^-+\eta'-\frac{1}{2}\eta\,\eta'\,\eta\Big)G_t\,\Upsilon^{-1}\,G_t^{-1}\\
\nn 
&&+\frac{i}{2}\,\kappa\,\eta\,G_t\,\Upsilon\,G_t^{-1}\Big(\frac{i}{2}[\Sigma_-,\eta]x'^-+\eta'-\frac{1}{2}\eta\,\eta'\,\eta\Big)G_t\,\Upsilon^{-1}\,G_t^{-1}\,\eta\Big\}+\mathcal{O}(\eta^6).
\eea
Where the leading order term is the quadratic kinetic term and the higher order terms are just the
function $\widetilde{\Phi}(\eta)$ introduced earlier, which together with
its self interaction terms constitute
the fermionic shift.

Since we do a perturbative analysis up to quartic order, the presence of
quadratic fermionic terms in the above expressions imply that we need $\PI_\pm$
to quadratic order only\footnote{This is only true for the fermionic kinetic
term. In the full Lagrangian $\PI_-$ is needed to quartic order.}, from (\ref{pi})
we find
\bea 
\label{piquadratic}
&& g\,\PI_-=\frac{i}{4}\Big(p_y^2+\frac{1}{2}Tr(P_z\cdot P_z)+4\,p_{\dot{a}}\,\bar{p}^{\dot{a}}+\omega'_{\dot{a}}\,(\bar{\omega}')^{\dot{a}}
+y'^2+\frac{1}{2}Tr(Z'\cdot Z')\Big), \\ \nn
&& g\,\PI_+=\frac{i}{4}+\frac{i}{16}\Big(y^2-\frac{1}{2}Tr(Z\cdot Z)+\frac{1}{4}\omega_{\dot{a}}\,\bar{\omega}^{\dot{a}}\Big)
+\frac{1}{4}\Big(\bar{\Psi}'\,\gamma^1\,\Psi-\bar{\Psi}\,\gamma^1\,\Psi'-\frac{i}{2}\bar{\Psi}\cdot\Psi\Big).
\eea
Combining the solutions for $\PI_\pm$ and the transverse components of
$\PI$ in (\ref{pitrans}) one can solve for the fermionic shift (\ref{phi})
explicitly. As should
be clear, the explicit form in components
is quite complicated. Nevertheless, it is a straightforward task to obtain
the shift for each coordinates by inverting the expressions (\ref{fermlabel}).

To obtain the full shift that also removes the $Str\,\Phi_2\,\widetilde{\Phi}_2$
term, one need to isolate the $\dot{\eta}$ part and add this contribution
to (\ref{phi}). The terms from $Str\,\Phi_2\,\widetilde{\Phi}_2$ without
a $\dot{\eta}$ dependence will introduce corrections to $\PI_t$ which one
also need to determine explicitly. Having done all this, one can implement
the full shift in the original Hamiltonian, together with the corrections
to $\PI$, and determine the full mixing part of the shifted Hamiltonian.
Needless to say, all this will be a rather involved procedure and is beyond
the scope of this paper.
\end{appendix}

\end{document}